\documentclass[preprint,showpacs,titlepage,aps,prd,
tightenlines,amsmath,byrevtex,nofootinbib]{revtex4}
\usepackage{graphicx}

\def\lsim{\raise0.3ex\hbox{$\;<$\kern-0.75em\raise-1.1ex
\hbox{$\sim\;$}}}
\def\gsim{\raise0.3ex\hbox{$\;>$\kern-0.75em\raise-1.1ex
\hbox{$\sim\;$}}}

\begin{document}

\preprint{hep-ph/0701151}

\title{Determination of the Neutrino Mass Hierarchy via the Phase of the Disappearance Oscillation Probability with a Monochromatic $\bar{\nu}_e$ Source.}

\author{H.~Minakata$^{1}$}
\email{minakata@phys.metro-u.ac.jp}
\author{H.~Nunokawa$^{2}$}
\email{nunokawa@fis.puc-rio.br} 
\author{S.~J.~Parke$^{3}$}
\email{parke@fnal.gov} 
\author{R.~Zukanovich Funchal$^{4}$}
\email{zukanov@if.usp.br}
\affiliation{
$^1$Department of Physics, Tokyo Metropolitan University, 
Hachioji, Tokyo 192-0397, Japan \\
$^2$Departamento de F\'{\i}sica, Pontif{\'\i}cia Universidade Cat{\'o}lica 
do Rio de Janeiro, C. P. 38071, 22452-970, Rio de Janeiro, Brazil \\
$^3$Theoretical Physics Department,
Fermi National Accelerator Laboratory, 
P.\ O.\ Box 500, Batavia, IL 60510, USA \\
$^4$Instituto de F\'{\i}sica, Universidade de S\~ao Paulo, 
 C.\ P.\ 66.318, 05315-970 S\~ao Paulo, Brazil
}
\date{July 13, 2007}

\vglue 1.6cm

\begin{abstract}
The neutrino mass hierarchy can be determined, in principle, by
measuring a phase in the disappearance oscillation probability in
vacuum, without relying on the matter effect, using a single
channel. This phase is not the same for the normal and inverted
neutrino mass spectra. In this paper, we give a complete description
and physics understanding of the method.  The key feature of the
method is to detect advancement (normal) or retardation (inverted) of
the phase of the atmospheric-scale oscillation relative to the
solar-scale oscillation.
We then show that this method can be realized with the recently
proposed resonant $\bar{\nu}_{e}$ absorption reaction enhanced by
M\"ossbauer effect.  The unique feature of this setup is the
ultra-monochromaticity of the observed $\bar{\nu}_{e}$'s.  Because of
this feature the phase advancement or retardation of atmospheric-scale
neutrino oscillation is detectable after 20 or more oscillations if
the source and the target are made sufficiently compact in size.  A
quantitative estimate of the sensitivity to mass hierarchy resolution
is given.
%
We have also examined how a possible continuation of such
experiment can be carried out in order also to achieve high precision
(few \%) determination of the solar-scale oscillation parameters
$\Delta m^2_{21}$ and $\theta_{12}$.

\end{abstract}


\maketitle

\section{Introduction}
\label{sec:intro}

Among the remaining unknowns in neutrino physics, resolving the
neutrino mass hierarchy may be one of the key importance to our
understanding of physics of neutrino masses and the lepton flavor
mixing \cite{review}.  Given the significance of the problem, it is
natural that many methods are proposed for determining the mass
hierarchy.
The most conventional one would be to do neutrino and antineutrino
comparison in long-baseline (LBL) accelerator based 
oscillation experiments \cite{LBL_matter1,LBL_matter2}.  
Because of the difference in how the
matter effect interferes with the vacuum oscillation effect it can
signal the sign of $\Delta m^2_{31}$, and hence the mass hierarchy.
Other less conventional ideas include the methods using atmospheric
neutrinos \cite{atm-method}, $\nu_{e}$ and $\nu_{\mu}$
disappearance channels \cite{GJK,NPZ,MNPZ}, supernova neutrinos
\cite{SN-method1,SN-method2}, and neutrino-less double beta decay
\cite{double-beta}.  These lists of references cannot be complete, and
some of the other relevant references are quoted therein.

In this paper, we discuss yet another way of determining the neutrino
mass hierarchy by a neutrino oscillation disappearance experiment
using {\it a single neutrino flavor} in vacuum whereas in previous 
works~\cite{NPZ,MNPZ} the method required the use of two different 
disappearance channels close to the first oscillation maximum.
The possibility that the neutrino survival probability in vacuum can
be sensitive to the neutrino mass hierarchy was first suggested in
\cite{petcov} using reactor neutrinos in the context of high-$\Delta
m^2$ large mixing angle (LMA) solar neutrino solution.  It was also
pursued recently in the Hanohano project using a Fourier transform
technique \cite{hanohano}.  We present here a complete formulation and
physics understanding of the new method to determine the neutrino mass
hierarchy.

This method utilizes the feature that the atmospheric-$\Delta m^2$
scale oscillation ``interferes'' with the solar-$\Delta m^2$ scale
oscillation in a different way depending upon the mass hierarchy.  We
first carefully re-formulate the difference as an advancement or
retardation of the phase of the atmospheric-scale neutrino oscillation
relative to the solar-scale oscillation; For the $\nu_e$ and
$\bar{\nu}_e$ disappearance probabilities, the atmospheric oscillation
is continuously advanced (retarded) for the normal (inverted)
hierarchy.
Since the core of the problem exists in extracting long wavelength
solar oscillation out of the two short wavelength atmospheric scale
oscillations due to $\Delta m^2_{32}$ and $\Delta m^2_{31}$, it is the
critical matter to identify which quantity to be held fixed in
defining the mass hierarchy reversal from the normal to the inverted,
or vice versa.  To our understanding, the only known appropriate
quantity to hold is the effective neutrino mass squared difference
$\Delta m^2_{\text{ee}}$ (see Eq.(\ref{eqn:dmsqee})), the one which is
actually observed in $\nu_{e}$ (or $\bar{\nu}_{e}$) disappearance
measurement at the first few oscillation maxima and/or minima
\cite{NPZ}.  We find it important to have the proper formalism because
otherwise the sensitivity to the mass hierarchy resolution can be
significantly overestimated.

Then, we point out that an experiment using the resonant
$\bar{\nu}_{e}$ absorption reaction \cite{mikaelyan} enhanced by
M\"ossbauer effect recently proposed by Raghavan \cite{Raghavan} gives
the first concrete realization of the idea.  (See \cite{early} for
earlier proposals of M\"ossbauer enhanced experiments.)
Because of the monochromaticity of the $\bar{\nu}_{e}$ beam from 
the bound state beta decay \cite{bahcall}, 
the phase advancement or retardation of atmospheric-scale neutrino 
oscillation is detectable after 20 or more oscillations if the source 
and the target are made sufficiently compact in size.
The unique feature of this method is the use of this ultra-monochromatic 
neutrino source thereby sidestepping the very high energy resolution
requirement requested with reactor or accelerator neutrinos.

A concrete plan for the experiment may be as follows; In the first
stage the atmospheric $\Delta m^2_{\text{ee}}$ must be accurately
measured in a way described in \cite{mina-uchi} by a measurement at
$\sim$10 m baseline. As we will see, the precision of the measurement
must reach sub-percent level in order for this method to work.  Then,
in the second stage relative phase advancement or retardation of the
atmospheric-scale neutrino oscillation must be detected by moving the
detector to $\sim$350 m, this is just beyond the solar-scale
oscillation distance for 18.6 keV neutrinos.  Based on this plan we
give a detailed estimate of the sensitivity to the resolution of the
mass hierarchy.

Upon finishing the above measurement, it is natural to think about the
continuation of the experiment by moving the detector to distances
tuned for precision measurement of the oscillation parameters relevant
for solar neutrinos, $\Delta m^2_{21}$ and $\theta_{12}$.  We also
give a detailed estimate of the accuracy for the determination of
these parameters. It is notable that a resonant $\bar \nu_e$
absorption experiment could allow to make precise measurements of {\it
all mixing parameters} that appear in the $\nu_e \to \nu_e$
oscillation channel: $\theta_{13}$, $\theta_{12}$, $\Delta m^2_{21}$, the 
effective atmospheric $\Delta m^2_{\rm ee}$ and the neutrino mass hierarchy.

In Sec.~\ref{sec:how}, we first describe the essence of the method in
an explanatory way.  We then define our theoretical machinery using
the effective $\Delta m^2_{\text{ee}}$.  We also clarify the basic
requirement for the measurement.
In Sec.~\ref{sec:exp}, we discuss the concrete set up of the experiment 
and describe our statistical method for the analysis. 
We then give the analysis results on sensitivity to the mass hierarchy 
resolution. 
In Sec.~\ref{sec:solar}, we describe the analysis for the determination of
the solar oscillation parameters.
In Sec.~\ref{sec:conc}, we finish by giving our concluding remarks. 
In Appendix ~\ref{appendixA}, we give a different 
interpretation of the phenomena in terms of the effective 
mass squared differences whereas in Appendix~\ref{appendixB}, 
we discuss how this method can be applied to 
the $\nu_\mu$ disappearance channel. 
This possibility has been discussed in \cite{gouvea-winter} however 
the baseline required is $L \simeq11000 \left( E / 0.6 \text{ GeV} \right)$ km.

\section{How can $\nu_{e}$ disappearance measurement alone 
signal the neutrino mass hierarchy?}
\label{sec:how}

The vacuum $\nu_e$ survival probability 
%
using 
$\Delta_{ij} \equiv \Delta m^2_{ij} L/4E$ ($\Delta m^2_{ij}\equiv m^2_i-m^2_j$)
as shorthand for the kinematical phase,
can be written without any approximation as
\begin{eqnarray}
P(\nu_e \rightarrow \nu_e) & = & 1- 
4\vert U_{e3}\vert^2 \vert 
U_{e1}\vert^2 \sin^2 \Delta_{31} -
4 \vert U_{e3}\vert^2 \vert U_{e2}\vert^2 \sin^2 \Delta_{32}  -
P_{\odot}, 
\nonumber \\
&=& 
1-\frac{1}{2} \sin^2 2 \theta_{13} 
\left[
1 - \left( c^2_{12} \cos 2 \Delta_{31} + s^2_{12} \cos 2 \Delta_{32} \right)  
\right] - P_{\odot}, 
\label{Pee}
\end{eqnarray}
where the usual parameterization of the mixing matrix~\cite{PDG} is used 
with the abbreviations $c_{ij}\equiv \cos\theta_{ij}$ and $s_{ij}\equiv
\sin\theta_{ij}$. Here $E$ is the neutrino energy and $L$ is the baseline 
distance. $P_{\odot}$, as defined by,
\begin{eqnarray}
P_{\odot} \equiv 
4\vert U_{e2}\vert^2 \vert U_{e1}\vert^2 \sin^2 
\Delta_{21} = 
\sin^2 2 \theta_{12} c^4_{13}\sin^2 \Delta_{21},
\end{eqnarray}
is the oscillation probability associated with the solar $\Delta m^2$
scale. In this paper,  
we assume the standard three neutrino framework with CPT 
conservation (except for where we discuss possible CPT violation), 
which implies that Eq.~(\ref{Pee}) is also valid for $\bar\nu_e$.

\begin{figure}[b]
\vglue 0.2cm
\begin{center}
\includegraphics[width=0.49\textwidth]{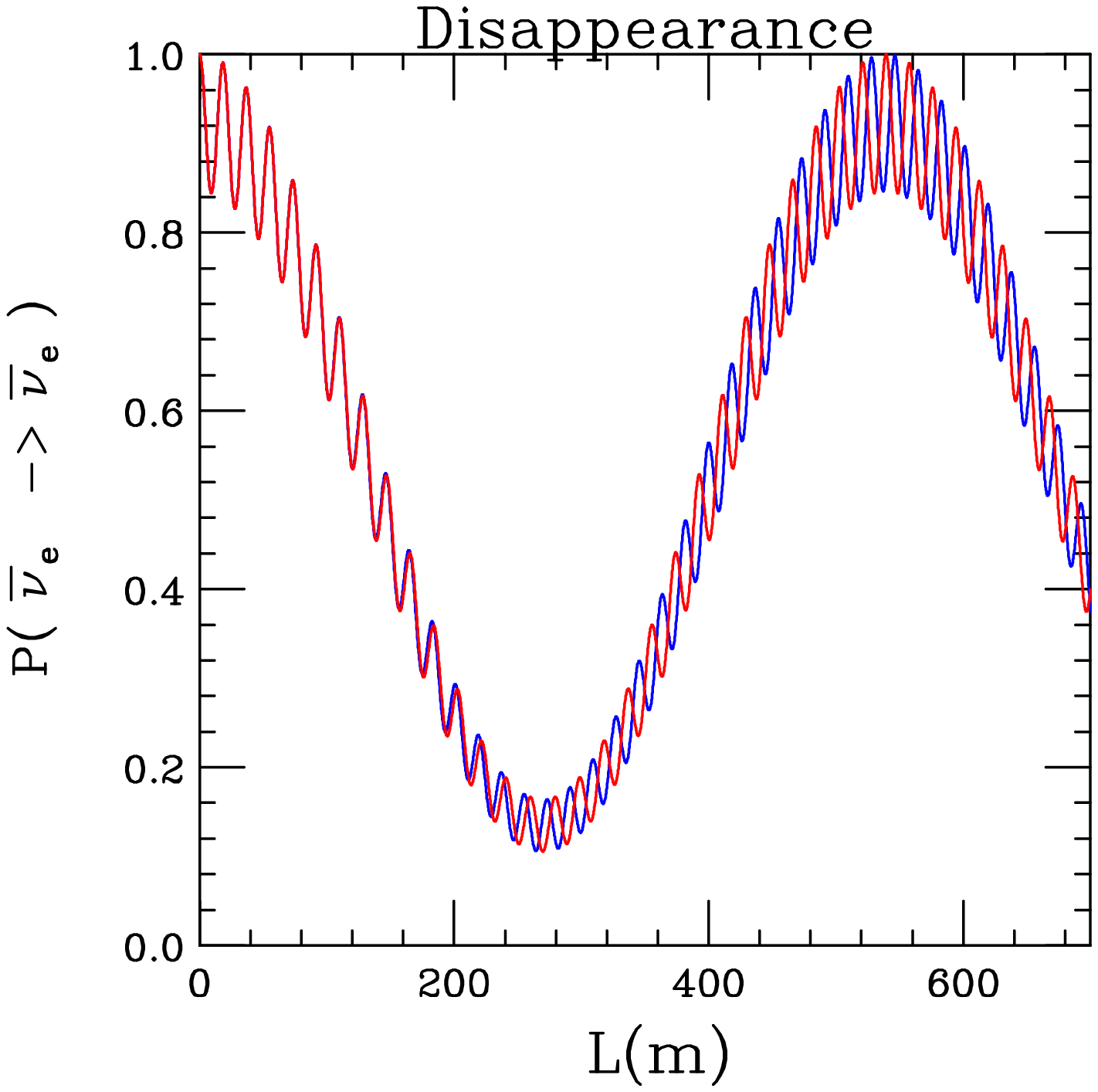}
\includegraphics[width=0.49\textwidth]{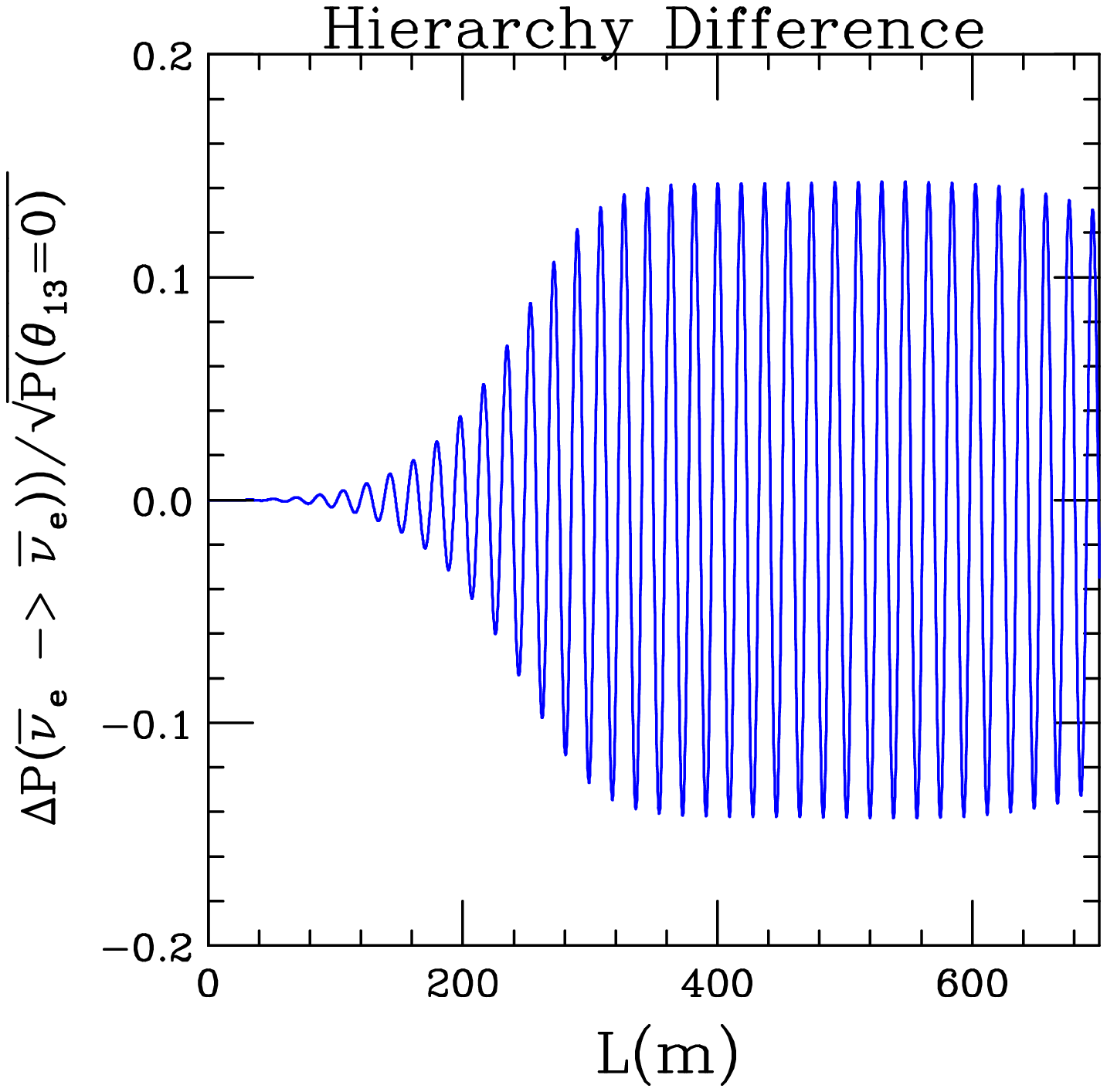}
\end{center}
\vglue -0.8cm
\caption{In the left panel, antineutrino survival probability
  $P(\bar{\nu}_e \rightarrow \bar{\nu}_e)$ is plotted as a function of
  $L$. The blue and the red curves are for the normal and the inverted
  mass hierarchies, respectively.  In the right panel, $\Delta P /
  \sqrt{P(\theta_{13}=0)}$, which is defined in the text, is shown. 
  Here the neutrino energy was fixed to 18.6 keV. This choice comes from 
  the experiment we will discuss later in this paper.}
\label{Rio-SP}
\end{figure}

The neutrino mass hierarchy can be determined if one knows which
kinematical phase, $\vert \Delta_{31} \vert $ or $ \vert \Delta_{32}
\vert $, is larger.  Namely, if $\vert \Delta_{31} \vert > \vert
\Delta_{32} \vert $ the normal hierarchy is the case, and if $\vert
\Delta_{31} \vert < \vert \Delta_{32} \vert $ the inverted one.
How we can know which is larger? 
When two waves with slight different frequencies travel together there
arises the phenomenon of beating. The amplitude of superposed waves
beats with the frequency corresponding to the difference of two
frequencies.  For $\nu_e$ disappearance probability, Eq.~(\ref{Pee}),
we have beating because the frequency of the two waves in
(\ref{Pee}) with $\vert \Delta_{31} \vert $ and $ \vert \Delta_{32}
\vert $ differ by the frequency associated with the $\vert \Delta_{21}
\vert $ characterized by the solar $\Delta m^2_{21}$ scale, which is
smaller by a factor of $\simeq 30$ compared to the atmospheric $\vert
\Delta_{31} \vert $ scale.
One can decompose the superposed waves into the beating low frequency
wave and the high frequency wiggles within the beat.  Here, unlike the
case addressed in most textbooks of basic physics, the amplitude of
the two frequencies is different.  Therefore, to tell which of the
high frequencies has a larger amplitude, one has to determine whether
the phase of the high frequency wiggles advances or retards as the
wave evolves.
This phase advancement or retardation can tell us whether the neutrino
mass hierarchy is normal or inverted using only the disappearance
measurement in a single channel, $\nu_e \rightarrow \nu_e$ in vacuum.

To demonstrate that it is possible, we plot in the left panel of
Fig.~\ref{Rio-SP} the oscillation probability $P(\nu_e
\rightarrow \nu_e)$ as a function of the distance $L$ from a
source.  As one can clearly see in the figure, the case of normal
hierarchy ($\Delta m^2_{31} > 0$, blue curve) starts to be
distinguished from the case of inverted hierarchy ($\Delta m^2_{31} <
0$, red curve) at around $L \sim 300$ m because of the advancement of
the normal hierarchy's high frequency wiggles and the retardation of
the inverted hierarchy's high frequency wiggles, i.e. the phase of the
high frequency wiggles.
What we mean by ``beating'' may be clearer in the left panel of 
Fig.~\ref{fig:LoverE} where the solar term is switched off.

To see the difference between the oscillation probabilities of the
normal and the inverted hierarchies more clearly, we define the
quantity
\begin{eqnarray}
\Delta P(\nu_e \rightarrow \nu_e) & \equiv &
P(\nu_e \rightarrow \nu_e; \Delta m^2_{31} > 0) - 
P(\nu_e \rightarrow \nu_e; \Delta m^2_{31} < 0).
\label{DeltaP}
\end{eqnarray}
Where the hierarchy flip is defined so that $\vert \Delta P \vert$ 
is minimized for short baseline measurements. 
(See Sec.~\ref{subsec:theo} for the precise definition of the hierarchy flip.) 
We normalize $ \Delta P$ by dividing by $\sqrt{P(\theta_{13}=0)}$ since
$P(\theta_{13}=0)$ is the ``background'' for seeing $\theta_{13}$
effects in $\nu_e$ disappearance.  The quantity $ \Delta P /
\sqrt{P(\theta_{13}=0)}$ is plotted in the right panel of
Fig.~\ref{Rio-SP}.  The envelope of $\vert \Delta P \vert /
\sqrt{P(\theta_{13}=0)}$ slowly rises from 0 to a plateau value of
$\sin^2 2\theta_{13}$ at a distance about 25\% further from the source
than the first solar oscillation maximum ($\Delta_{21} = \pi / 2$
occurs when $L=280$ m) and remains reasonably constant at this value
beyond the first solar oscillation minimum ($\Delta_{21} = \pi$).
Clearly, even though the difference between the hierarchies is very
small at short distances it becomes as large as theoretically
possible, $\sin^2 2\theta_{13}$ (see below), at large distances.

\subsection{Theoretical framework}
\label{subsec:theo}

To decompose the superposed waves into the beat and wiggle,
one has to correctly define the high frequency mode. 
We choose to do this by defining a short distance effective $\Delta m^2$ in the $\nu_e$ 
disappearance channel, see~\cite{NPZ}. 
It is given by 
\begin{eqnarray}
  \Delta m^2_{\text{ee}} \equiv  \cos^2 \theta_{12} \vert \Delta m^2_{31}  \vert + \sin^2 \theta_{12}  
  \vert \Delta m^2_{32} \vert .
\label{eqn:dmsqee}
\end{eqnarray}
This is the only linear combination of $ \Delta m^2_{31}$ and 
$ \Delta m^2_{32}$ that can be determined by a short distance 
(over the first few atmospheric oscillations) 
$\nu_e$ disappearance experiment.
$\Delta m^2_{\text{ee}}$ is defined to be positive for both hierarchies. 
The choice of the atmospheric $\Delta m^2$ is crucial; 
Extraction of the low frequency mode can be done correctly only if 
the high frequency mode is identified in an appropriate way. 
See Sec.~\ref{reversal} for more about it. 
%

Now, we rewrite $P(\nu_e \rightarrow \nu_e)$ without any
approximation, using the positive and measurable variables
$\Delta_{\text{ee}} \equiv \Delta m^2_{\text{ee}} L/4E$ 
and $\Delta_{21}$ and the phase $\phi_\odot$ (which will 
be given below), as follows:
\begin{eqnarray}
P(\nu_e \rightarrow \nu_e)& = & 1-
\frac{1}{2} \sin^2 2\theta_{13} \left[ 1-
\sqrt{1-\sin^2 2 \theta_{12} \sin^2 \Delta_{21}}
~\cos (2 \Delta_{\text{ee}} \pm \phi_\odot) \right]  - P_{\odot},
\label{eqn:Pdis}
\end{eqnarray}
where the positive (negative) {\it sign} in front of the phase 
$\phi_\odot$ corresponds to the normal (inverted) mass hierarchy; 
If $\Delta m^2_{\text{ee}} $ is precisely determined, the only 
effect of the hierarchy manifest itself as this {\it sign}.
%
Clearly, the amplitude of the $\sin^2 2\theta_{13}$ oscillations is
modulated by $\sqrt{ 1-\sin^2 2 \theta_{12} \sin^2 \Delta_{21}}$
which varies from 1 to $\cos 2\theta_{12}$ as the solar kinematic
phase, $\Delta_{21}$, changes from $0\,(\text{mod}\,\pi)$ 
to $\pi/2\, (\text{mod}\,\pi)$.
This amplitude modulation is the beat discussed in the previous
section.

\begin{figure}[bhtp]
\begin{center}
\hglue  -0.5cm
\includegraphics[width=0.46\textwidth]{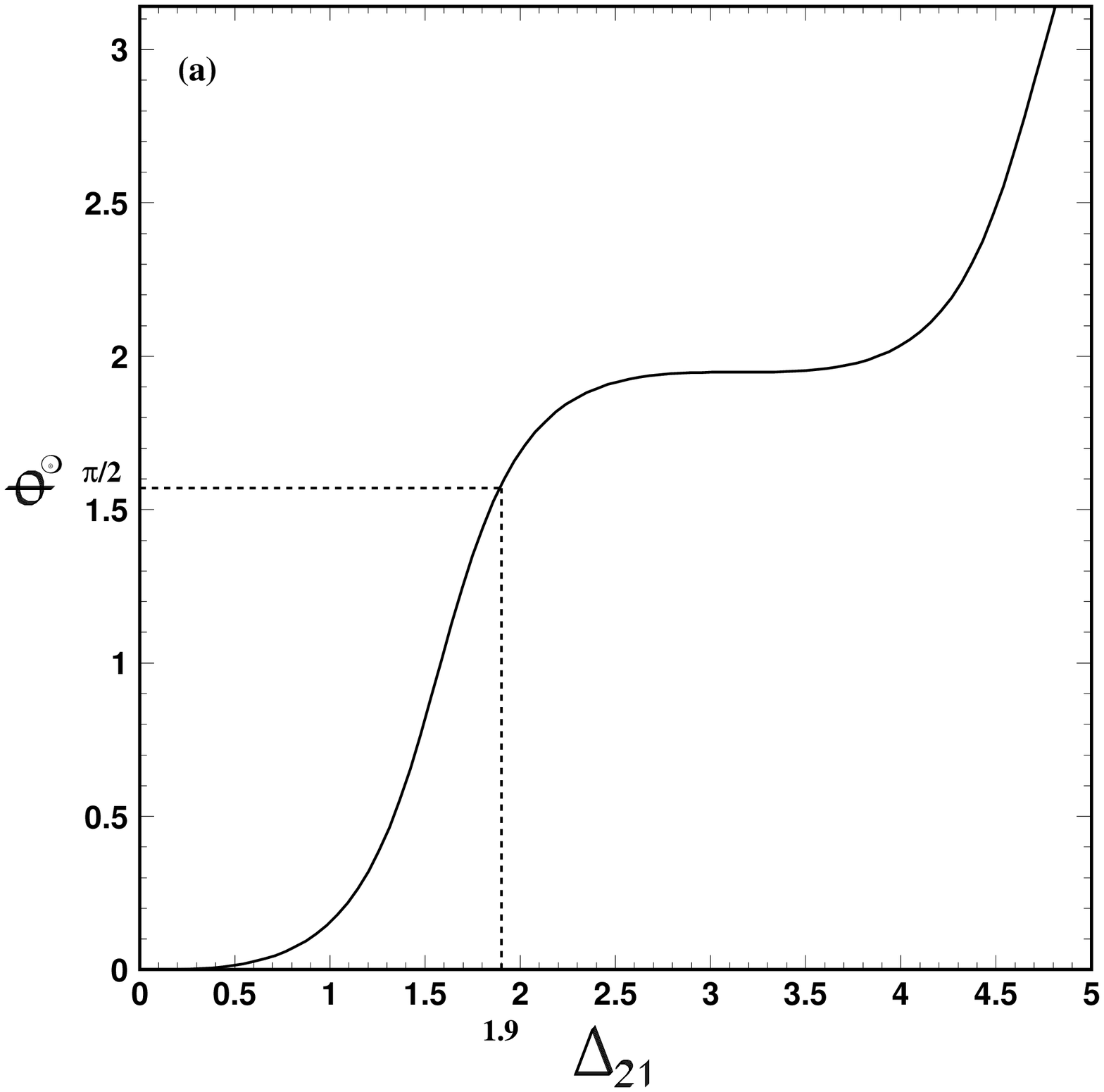}
\includegraphics[width=0.46\textwidth]{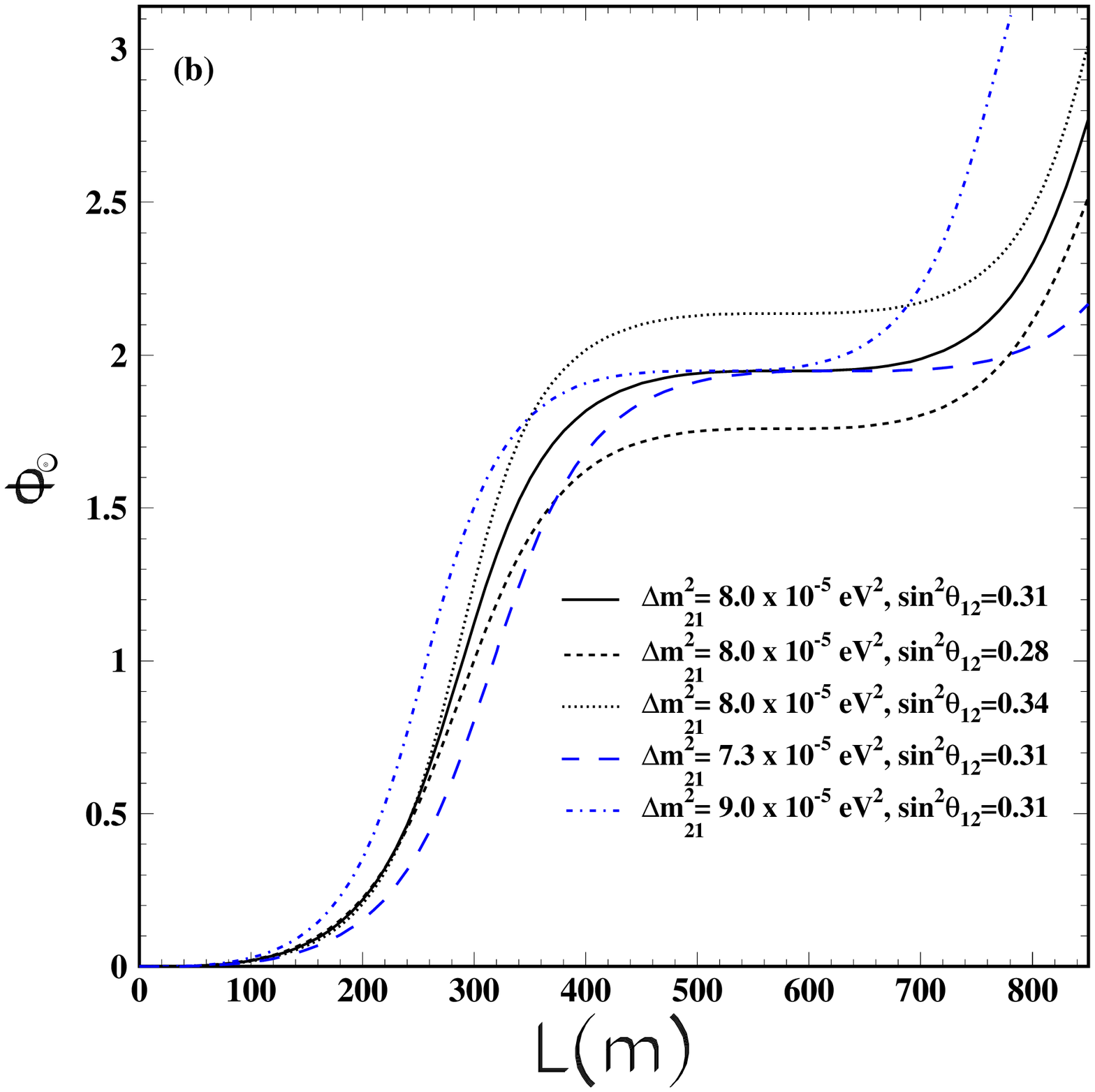}
\end{center}
\vglue -1.0cm
\caption{(a) $\phi_\odot$ as function of $\Delta_{21}$.  $\phi_\odot =
  \pi/2$ when $\Delta_{21}\approx 1.9$.(b) $\phi_\odot$ as function
  of $L$ for various choices of the solar parameters within the current 
allowed region. $\phi_\odot
  \approx \pi/2$ at about 350 m somewhat independent of the precise
  values for the solar parameters.}
\label{fig:phi}
\end{figure}

The phase $\phi_\odot$ in Eq.~(\ref{eqn:Pdis}) depends only 
on solar parameters,
$\Delta_{21}$ and $\theta_{12}$, and given as, 
\begin{eqnarray}
\sin \phi_\odot  & = & 
\frac{ c^2_{12} ~\sin ( 2 s^2_{12} \Delta_{21})
- s^2_{12}~ \sin ( 2 c^2_{12} \Delta_{21} ) }
{\sqrt{1-\sin^2 2 \theta_{12} \sin^2 \Delta_{21}}}, \nonumber \\
 \cos \phi_\odot & = & \frac{c^2_{12} ~\cos ( 2 s^2_{12} \Delta_{21})
 + s^2_{12} ~\cos ( 2 c^2_{12} \Delta_{21}) }
{ \sqrt{1-\sin^2 2 \theta_{12} \sin^2 \Delta_{21}}}.
\label{eqn:phi}
\end{eqnarray}
Or, alternatively
\begin{eqnarray}
  \phi_\odot  =  \arctan(\cos 2\theta_{12}\tan \Delta_{21})-\Delta_{21} \cos 2\theta_{12}. 
\label{eqn:phi2}
\end{eqnarray}
However, Eq.~(\ref{eqn:phi}) contains quadrant information not
available in Eq.~(\ref{eqn:phi2}).
In Fig.~\ref{fig:phi}, we show $\phi_\odot$ as a function of 
$\Delta_{21}$. Note that it is an odd, monotonically 
increasing function of $\Delta_{21}$. 

Due to the monotonically increasing nature of $\phi_\odot$ and the
$\pm$ sign in front of $\phi_\odot$, in Eq.~(\ref{eqn:Pdis}), for the
normal hierarchy the high frequency wiggles undergo a phase
advancement as the wave evolves, whereas for the inverted hierarchy
there is a phase retardation\footnote{Equivalently, one could interpret 
this phenomena as a change in the instantaneous effective 
$\Delta m^2_{\text{atm}}$. 
This interpretation is explored in Appendix A.}.
 It is easy to show that
 \begin{eqnarray}
   \phi_\odot(\Delta_{21}+\pi)   =  \phi_\odot(\Delta_{21}) + 2\pi \sin^2 \theta_{12},
 \end{eqnarray}
 i.e. the phase advancement (normal) or retardation (inverted) is
 $2\pi \sin^2 \theta_{12}$ for every $\pi$ increase of $\Delta_{21}$.
 Eqs.~(\ref{eqn:Pdis}) and (\ref{eqn:phi}) are the foundations of our
 investigation.

Some important observations are worth emphasizing immediately:

\begin{itemize}

\item 
Only $\Delta_{\text{ee}}$ varies at the atmospheric scale. Everything
  else, including phase $\phi_\odot$, varies at the solar scale. This
  is a useful distinction because these two scales differ by a factor of
  $\sim 30$.

\item 
The difference between probabilities corresponding to the two 
hierarchies (\ref{DeltaP}) is given by
\begin{eqnarray}
{\Delta P \over  \sqrt{P(\theta_{13}=0)}} & = &\sin^2 2\theta_{13} \sin 2\Delta_{\text{ee}} \sin \phi_\odot\, , 
\label{eq:pdif}
\end{eqnarray}
to leading order in $\theta_{13}$.  $\Delta P$ becomes visible when
the phase $\phi_\odot$ develops to order unity.  From
Fig.~\ref{fig:phi} this occurs at around the first solar oscillation
maximum, ($\Delta_{21}=\pi/2$), in agreement with the features
exhibited in Fig.~\ref{Rio-SP}. From (\ref{eq:pdif}) we also see that 
the normal and the inverted hierarchies are maximally out of phase when
$\phi_\odot =\pi/2$. This occurs when
\begin{eqnarray}
\cot \Delta_{21} \cot(\Delta_{21} \cos 2\theta_{12})
=-\cos 2 \theta_{12}\, .
\label{eqn:phieqpiover2}
\end{eqnarray}
The approximate solution to this equation, for $\theta_{12}$ around 30
degrees, is $\Delta_{21} \approx \frac{\pi}{2}+\cos 2\theta_{12}/(\pi
\sin^2 \theta_{12})$.  For $\sin^2 \theta_{12}=0.31$,
Eq.~(\ref{eqn:phieqpiover2}) is satisfied when $\Delta_{21} \approx
1.9$, i.e., just beyond the first solar oscillation maximum.

\item 
We notice that by doing transformation 
$\theta_{12} \rightarrow \pi/2 - \theta_{12}$ 
(or $s_{12} \leftrightarrow c_{12}$) in (\ref{eqn:phi}) flips the sign 
of  $\phi_{\odot}$. 
It means that the mass hierarchy is confused by the presence of 
the particular type of CPT violation, the subtle one discussed in \cite{andre}. 
It occurs because we use the antineutrino channel; 
If we are able to use the neutrino channel, the problem of confusion 
does not arise because we know that $\theta_{12} < \pi/4$ 
from the solar neutrino observation. 

\end{itemize}

\subsection{Experimental strategy and requirements to the accuracy of measurement} 
\label{subsec:req}

We now describe our experimental strategy for determining the
hierarchy.
First one needs to make a very precise measurement of $\Delta
m^2_{\text{ee}}$ at short distances from the source, 
at around half the atmospheric oscillation length, $L_0/2$
where 
\begin{eqnarray}
L_0 \equiv \frac{4 \pi E}{\Delta m^2_{\text{ee}} } 
= 18.4 \, 
\left[ \frac{ E }{18.6 \text{ keV}} \right] 
\left[ \frac{ \Delta m^2_{\text{ee}} }
{ 2.5 \times 10^{-3} \text{eV}^2} \right]^{-1} \text{m}\, .
\end{eqnarray}
Then, one must determine the sign in front of $\phi_\odot$ 
where the phase difference between normal and
inverted hierarchy is maximal, i.e. $\Delta_{21} \approx 2 $ as 
can be seen in Fig.~\ref{fig:phi}. 
With $\Delta_{\text{ee}} / \Delta_{21} \simeq 30$, it can be translated 
into the distance $L \approx [(2 \times 30) / \pi] L_0 \simeq 20 \, L_0$ 
away from the source.

How well do we need to know $\Delta m^2_{\text{ee}}$?  
If we do not know the value of $\Delta m^2_{\text{ee}}$
with sufficient precision, one can adjust $\Delta m^2_{\text{ee}}$, 
depending upon the normal (NH) and the inverted (IH) hierarchies, 
so that the relation 
\begin{eqnarray}
2 \Delta_{\text{ee}}\vert_{\text{NH}} + \phi_\odot = 
2 \Delta_{\text{ee}}\vert_{\text{IH}} - \phi_\odot 
\label{eqn:confusion}
\end{eqnarray}
holds  at some $L/E$. Then,  from Eq.~(\ref{eqn:Pdis})
the hierarchy is completely confused as we will explicitly 
see in Sec.~\ref{setup}.
It can be translated into 
\begin{eqnarray}
\frac{ \Delta m^2_{\text{ee}}\vert_{\text{IH}} - 
\Delta m^2_{\text{ee}}\vert_{\text{NH}} }
{\Delta m^2_{\text{ee}} } 
\left( \frac{\pi L}{L_0} \right) = 
 \phi_\odot
\label{eqn:confusion2}
\end{eqnarray}
which leads to the requirement for the accuracy of measurement of 
$\Delta m^2_{\text{ee}}$ as 
$\delta ( \Delta m^2_{\text{ee}} ) / \Delta m^2_{\text{ee}} \lsim 
2L_0 / \pi L \simeq 3$\%.  
This is the critical value at which the hierarchy is completely confused, 
and if we want to do better, e.g., at least a factor of 3 times 
better than 3\%, 
the first requirement is $\simeq$1\% measurement of $\Delta m^2_{\text{ee}}$.

\begin{figure}[tbhp]
\begin{center}
\hglue  -0.5cm
\includegraphics[width=0.33\textwidth]{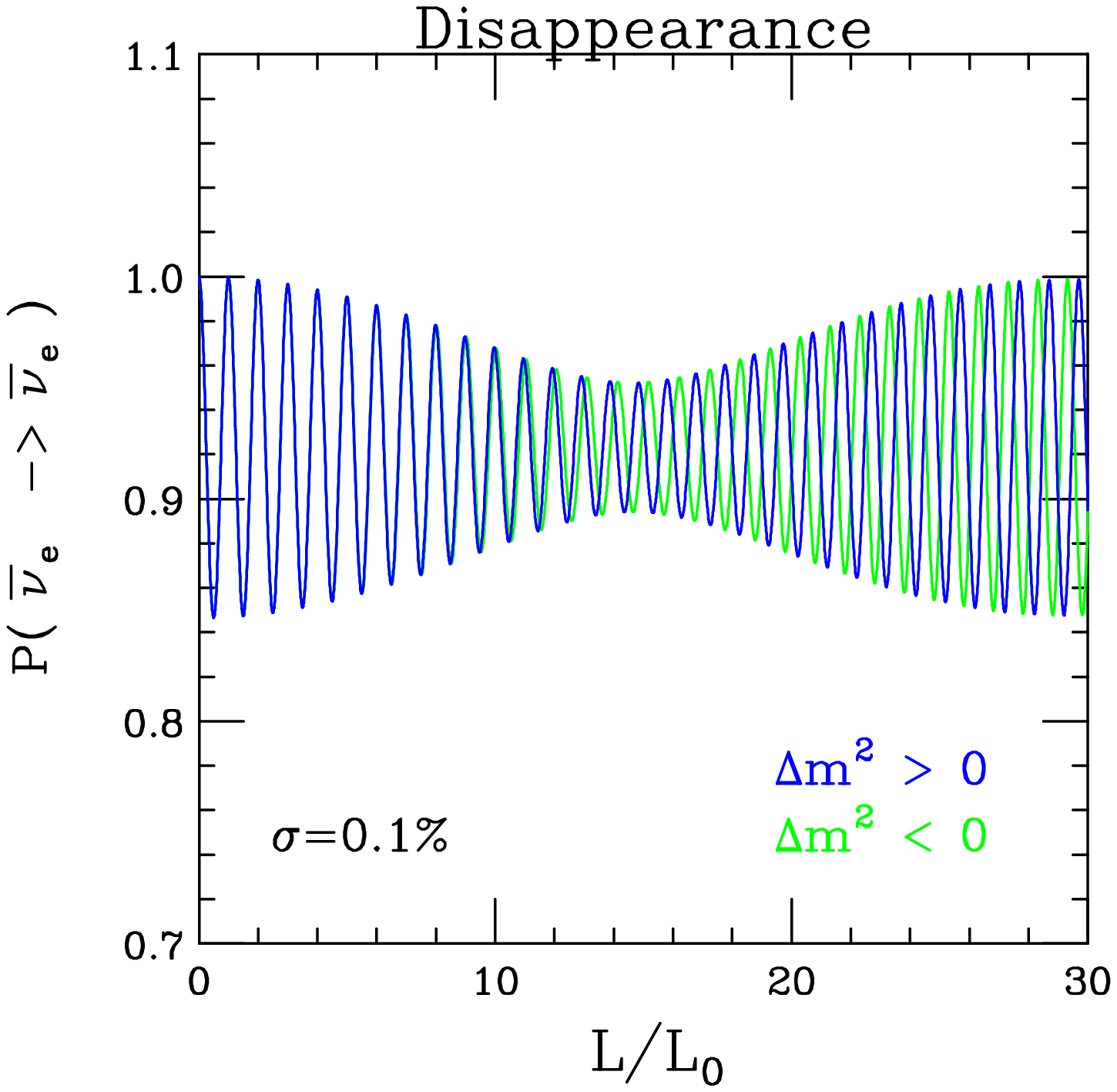}
\includegraphics[width=0.33\textwidth]{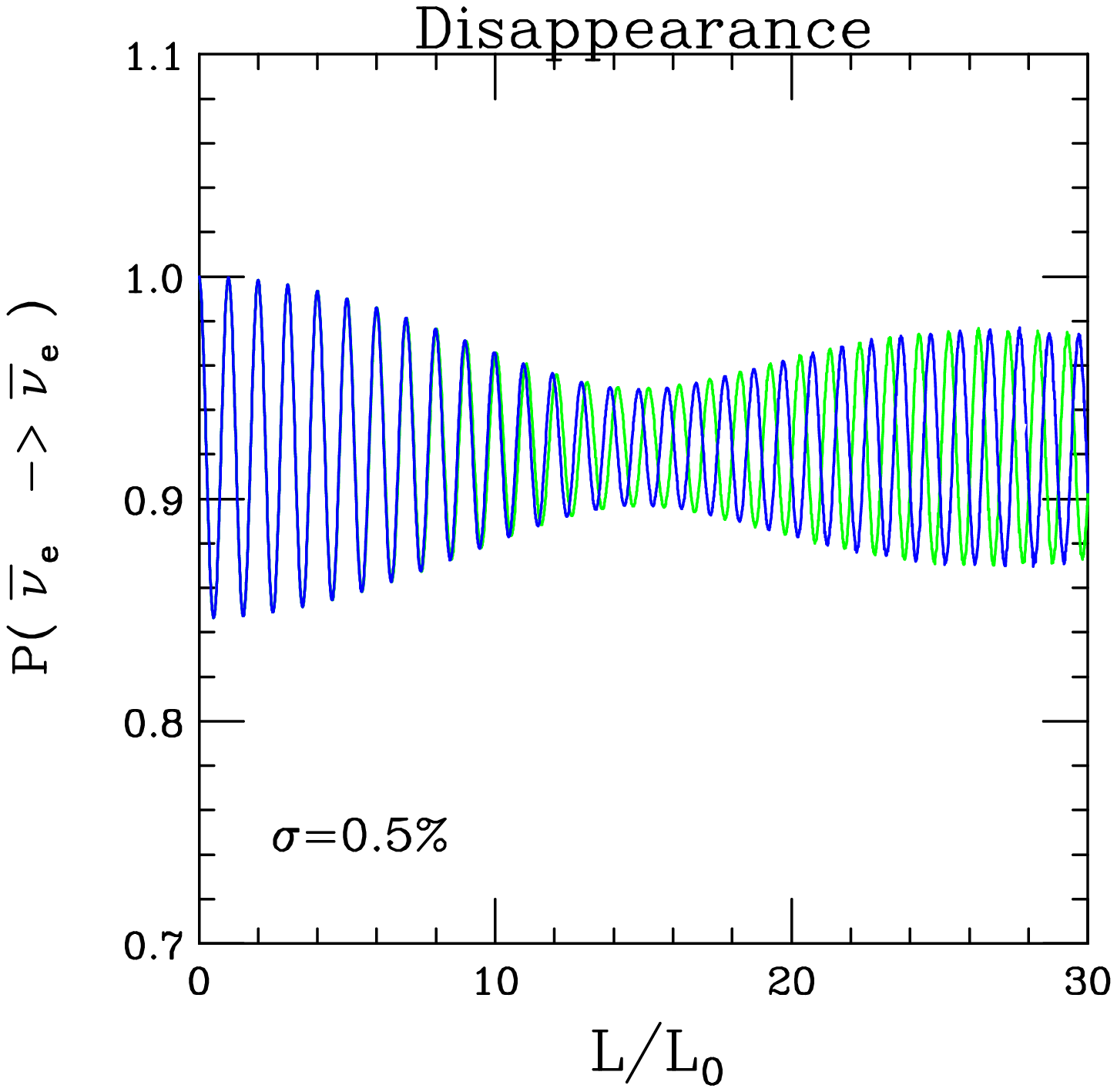}
\includegraphics[width=0.33\textwidth]{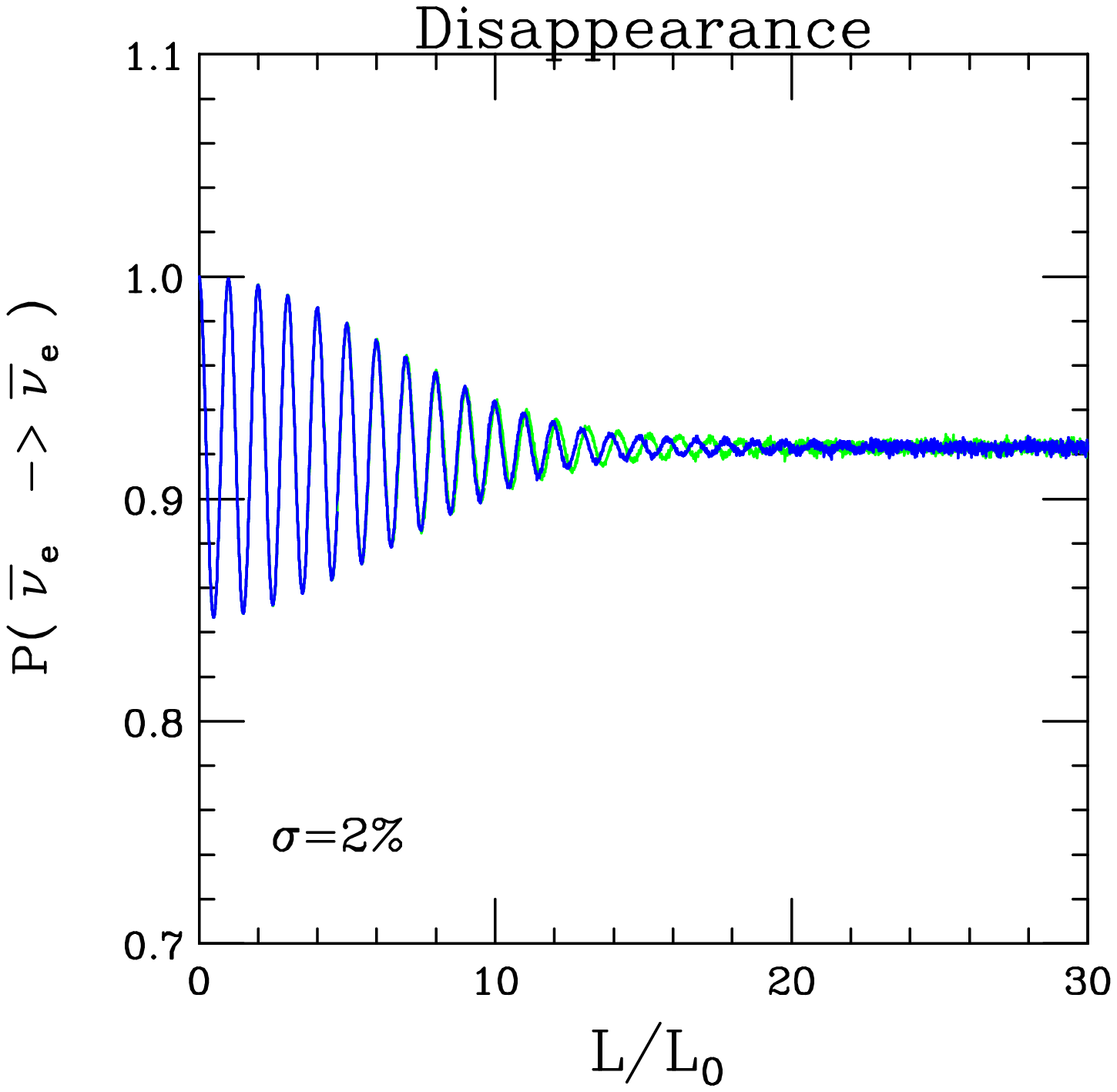}
\end{center}
\vglue -0.5cm
\caption{The averaged $P(\nu_e \rightarrow \nu_e)$ versus distance
  measured in number of oscillation lengths, $L_0$. 
  To help with visualization $P_\odot=\sin^2 2 \theta_{12}
  \cos^4 \theta_{13} \sin^2 \Delta_{21}$ has been set to zero for
  these figures. (a) 0.1\% Gaussian resolution on $L/E$.  (b) 0.5\%
  resolution on $L/E$. (c) 2\% resolution on $L/E$. In (b) the
  oscillations are reduced at 20 or so oscillations out, but still
  observable. Whereas in (c) the oscillations are averaged out by 15 $L_0$.
}
\label{fig:LoverE}
\end{figure}

At this point the astute reader will be concerned that we are
attempting to determine a oscillation phase after 20 or so
oscillations.  Aren't the atmospheric oscillations completely washed
out long before 20 $L_0$?  To observe atmospheric oscillations
20 or so oscillations out, assuming coherence, 
%
demands that we can determine $L/E$ per event to high precision.  
Figure~\ref{fig:LoverE} shows how the oscillations are
reduced as the resolution on $L/E$ is increased and that a precision
better than 1\% is required to see oscillation at $\sim$ 20  
oscillation lengths from the source.  

In summary, to measure the sign of the phase and determine the
hierarchy we have {\it two} important initial requirements:
\begin{itemize}

\item The resolution on $ \Delta m^2_{\text{ee}}$ measured in a short
baseline experiment ($\sim L_0/2$ from the source) must be $\simeq$1\%
or less.

\item The resolution on $L/E$ for the long baseline experiment ($\sim$
20 $L_0$ from the source) must be $\simeq$1\% or less.

\end{itemize}
Usually, the neutrino energy cannot be determined experimentally with
this accuracy, therefore, these preconditions are very difficult to
meet, if not impossible.

\subsection{Mass hierarchy reversal and comments on the reactor neutrino method}
\label{reversal}

We have emphasized in the foregoing discussions the importance of
identifying the quantity to be held fixed to define the mass hierarchy
reversal, and proposed $\Delta m^2_{\text{ee}}$ as the solution.  In
this subsection we want to elaborate this point and clarify how the
difference between the normal and the inverted hierarchies can be made
artificially larger by choosing different variables to define the
hierarchy reversal.

\begin{figure}[tbhp]
\begin{center}
\hglue  -0.5cm
\includegraphics[width=0.33\textwidth]{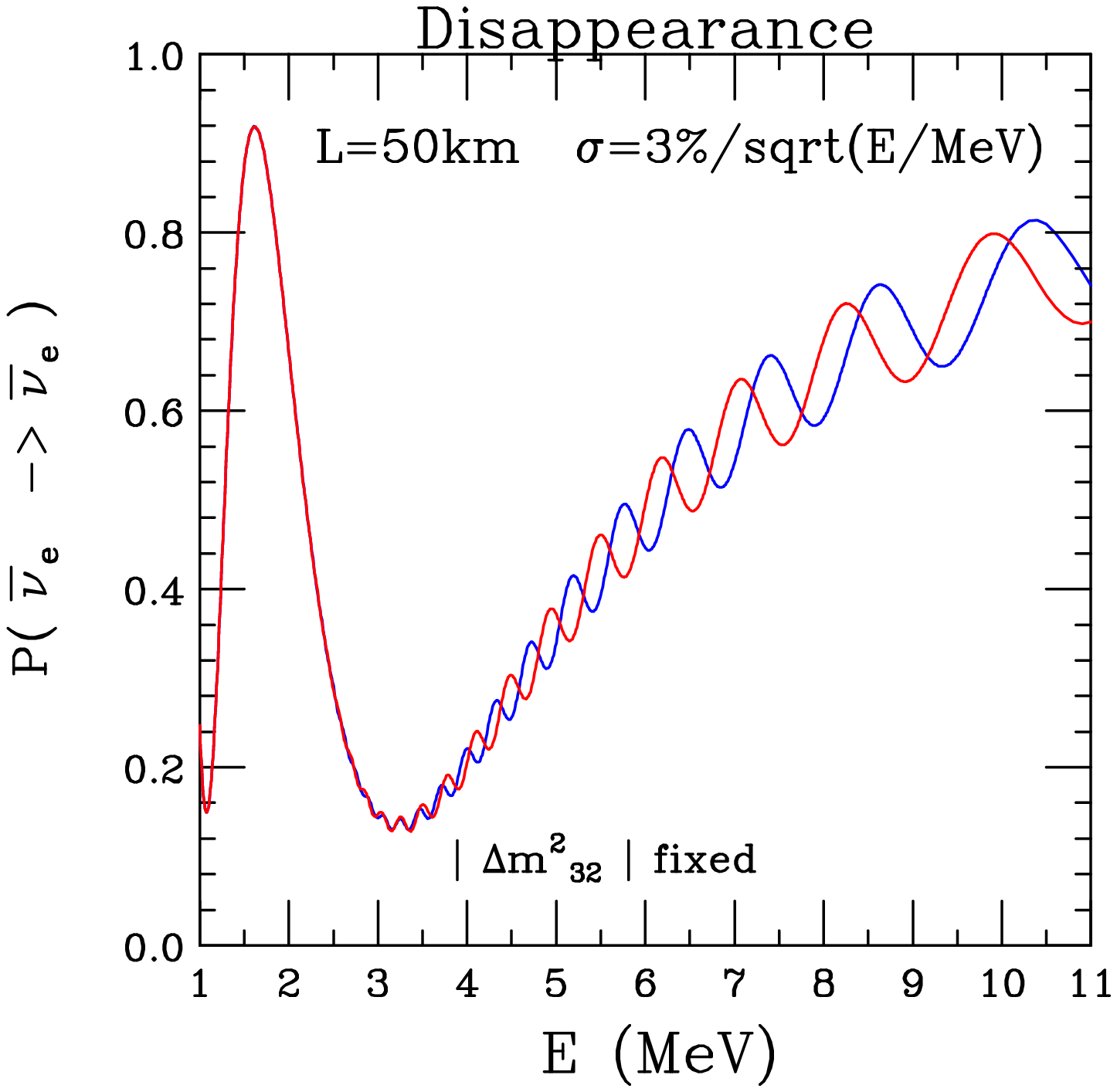}
\includegraphics[width=0.33\textwidth]{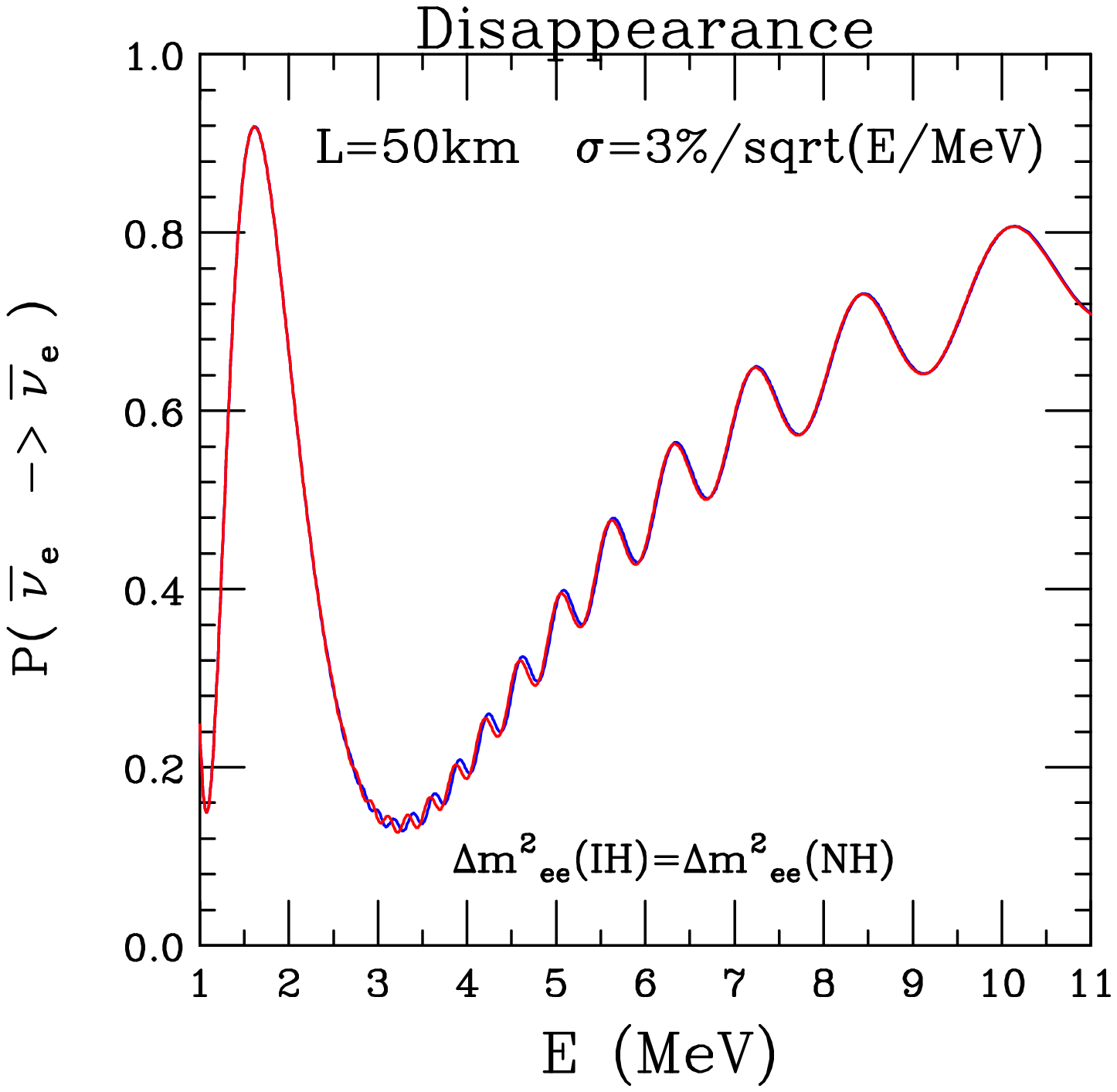}
\includegraphics[width=0.33\textwidth]{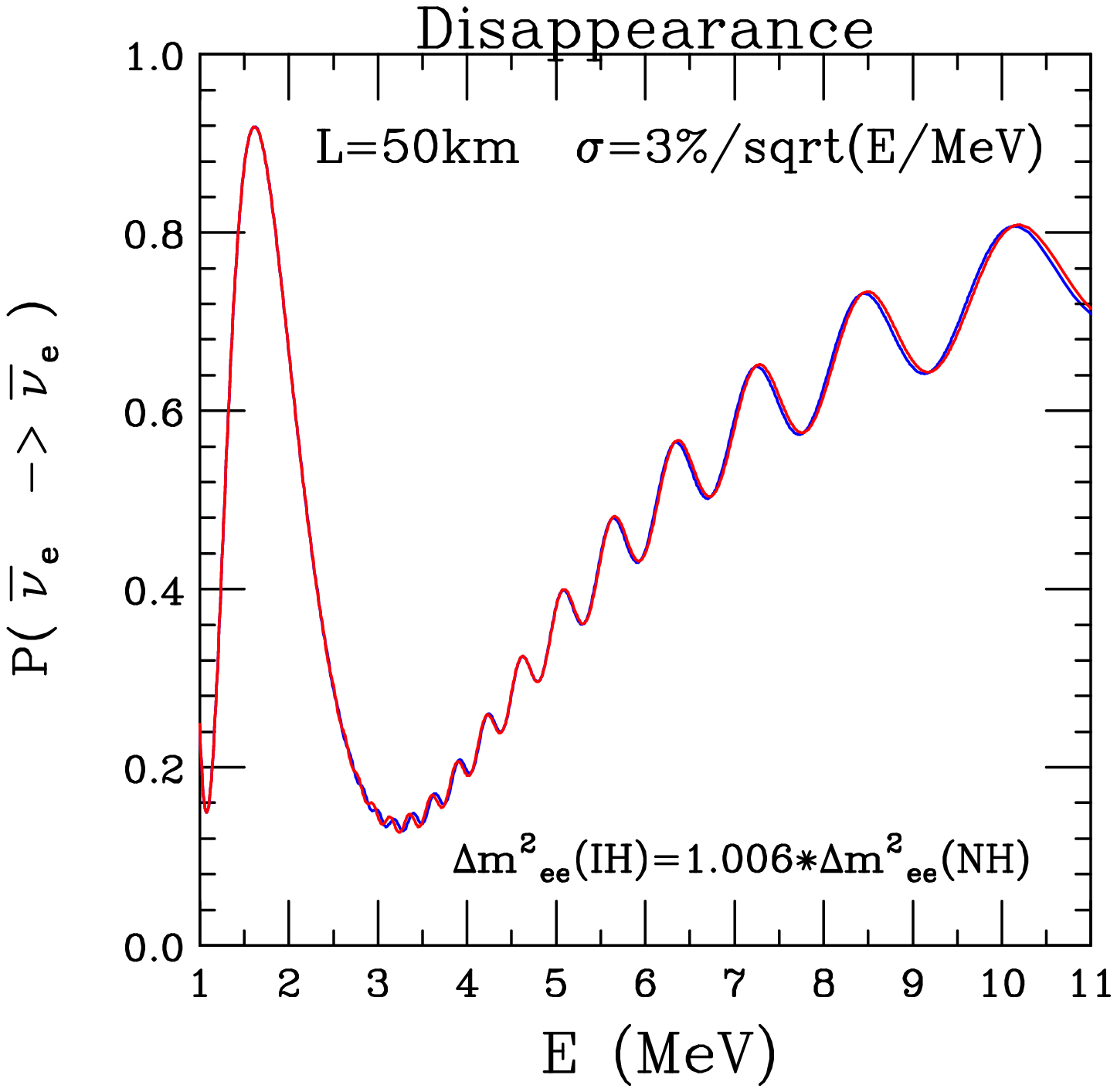}
\end{center}
\vglue -0.5cm
\caption{The $\bar{\nu}_e$ disappearance probability at 50 km from a 
reactor source as a function of the neutrino energy. The resolution 
on the energy is assumed to be 3\%/$\sqrt{E / \text{MeV} }$.
The blue (light gray) and the red (dark gray) curves are for the normal 
and the inverted mass hierarchies, respectively. 
In the left panel $|\Delta m^2_{32}|$ is held fixed when 
the hierarchy is flipped, whereas in the middle panel $\Delta m^2_{\text{ee}}$ 
is held fixed. In the right panel $\Delta m^2_{\text{ee}}$ for the 
inverted hierarchy is 0.6\% larger then for the normal hierarchy.
}
\label{fig:mh-reversal}
\end{figure}

In Fig.~\ref{fig:mh-reversal}, presented is the survival probability
of electron anti-neutrinos at a baseline of 50 km from a source which
is averaged over the uncertainty of energy 3\%/$\sqrt{E / \text{MeV}
}$ as a function of neutrino energy $E$.  It is shown in the left
panel of Fig.~\ref{fig:mh-reversal} that if we hold $\vert \Delta
m^2_{32} \vert $ fixed in reversing the hierarchy from the normal to
the inverted (as was done in \cite{petcov}) the difference between the
normal and the inverted hierarchies is clearly visible.
However, the obvious distinction seen in the left panel of
Fig.~\ref{fig:mh-reversal} disappears if we use $\Delta
m^2_{\text{ee}}$ to hold when switching between the hierarchies as
shown in the middle panel of Fig.~\ref{fig:mh-reversal}.  A simple
explanation for such a marked difference is that by holding $\vert
\Delta m^2_{32} \vert $ fixed in reversing the hierarchy from the
normal to the inverted one maps the neutrino mass spectrum into a
significantly different one, $\vert \Delta m^2_{31}\vert_{\text{IH}} =
\vert \Delta m^2_{31}\vert_{\text{NH}} - 2 \Delta m^2_{21}$.  (See
Appendix~\ref{appendixA} for more about it.)  This artificially
enhances the difference between the hierarchies, as demonstrated in
Fig.~\ref{fig:mh-reversal}.

Despite that there exist some discrepancies at energies around 4 MeV,
a peak energy of signal events in reactor experiments, the two
disappearance probabilities can be made very close to each other over
the energy range of 4 to 6 MeV if we allow the value of $\Delta
m^2_{\text{ee}}$ for the inverted hierarchy to increase by 0.6\%
relative to the value for the normal hierarchy, as can be seen in the
right panel of Fig.~\ref{fig:mh-reversal}.  The lesson we have learned
from this exercise is that unless we know $\Delta m^2_{\text{ee}}$ to
a sub-percent level, the hierarchy can be easily confused by a tiny
shift of the atmospheric $\Delta m^2$.

One might want to use neutrinos from reactors to determine the mass
hierarchy by the present method, as suggested by the authors of
\cite{petcov} . However, the experimental requirements are very severe
in order for this method to work. We believe that the three points
discussed above deserve attention if one want to make a reliable
estimate of sensitivity to the mass hierarchy resolution with reactor
neutrinos; (1) definition of hierarchy reversal, (2) implementation of
realistic energy resolution, and (3) robustness of the sensitivity to
mass hierarchy resolution against the uncertainty of $\Delta
m^2_{ee}$.

The recent Hanohano proposal, \cite{hanohano}, sidesteps some of these
requirements by using a sophisticated Fourier transform
technique. Detailed analysis around the atmospheric peak in Fourier
space can tells us whether $\vert \Delta m^2_{32} \vert$ is greater or
smaller than$\vert \Delta m^2_{31} \vert$, thus determining the
hierarchy. There is no need to explicitly deal with the definition of
hierarchy reversal. In this way, the two curves shown in the two
right panels of Fig.~\ref{fig:mh-reversal} can, in principle, be
distinguished. If this can be achieved in a realistic setting of the
experiment, it represents a remarkable advance for reactor neutrino
experiments, demonstrating the power of the Fourier transform
technique.

\section{Experimental setup, analysis method and results}
\label{sec:exp}

Most probably, the unique possibility to satisfy both requirements of
the previous section is to make use of monochromatic beam from the
bound-state beta decay $^{3}\mbox{H} \rightarrow \, ^{3}\mbox{He} +
\mbox{orbital e}^{-} + \bar{\nu}_{e}$ and detect it by the
time-reversed resonant absorption reaction $\bar{\nu}_{e} +
^{3}\mbox{He} + \mbox{orbital e}^{-} \rightarrow \, ^{3}\mbox{H}$, the
idea recently revived by Raghavan \cite{Raghavan}.
%
For the M\"ossbauer enhancement to work the method requires 
$\Delta E/E$ better than $10^{-16}$, 
a truly monochromatic neutrino beam of 18.6 keV. 
Thus, the energy spread of the beam in such an experiment is 
automatically much narrower than what is required. 


Having monochromaticity satisfied, the remaining 
requirement is on the baseline $L$. 
The short and long baseline measurement are to be done 
at the distances of $\sim$10 m and $20 L_0 \sim 350$ m 
from the source, respectively. 
Then, the requirement on $\Delta L / L$ of, say, 0.3\% implies 
that the size of the source and the target must be limited respectively to 
$\sim$3 cm and $\sim$1 m in short and long baseline measurement.  
The former places severe limits on the sizes of the source and the detector. 
Assuming this requirement can be met, 
it has been shown by analyzes with several concrete settings
that a sensitivity to $\Delta m^2_{\text{ee}}$ of $\simeq 0.3~(\sin^2
2\theta_{13} / 0.1)^{-1}$\% at 1$\sigma$ CL is
possible~\cite{mina-uchi}.  Therefore, the first requirement listed in 
Sec.~\ref{subsec:req} can be met if $\sin^2 2\theta_{13} \gsim$ 0.05.

\subsection{Experimental setup and how two-stage measurement works}
\label{setup}

We consider a possible experimental set up for determination of the
neutrino mass hierarchy based on a M\"ossbauer enhanced $\bar\nu_e$ 
absorption experiment~\cite{Raghavan}.  
By using the revised estimate of the cross section 
$\sigma_{\text{res}} = 0.3 \times 10^{-32} \text{cm}^2$~\cite{Raghavan} 
the event rate is given by 
\begin{eqnarray}
R_{\text{enhanced}} = 3 \times 10^{5}
\left(  \frac{S \cdot M_{T}}{1 \text{MCi} \cdot \text{100 g} }  \right) 
\left(  \frac{L}{10~\text{m} }  \right)^{-2} 
\text{day}^{-1}. 
\label{enhancedR}
\end{eqnarray}
Therefore, one obtains about one million (2500) events per 3 days  (10 days)
for 1 MCi source and 100 g $^3$He target at a baseline distance $L=10$ (350) m.
The experiment should be done in two stages; In the first stage,
$\Delta m^2_{\text{ee}}$ and $\sin^2 2\theta_{13}$ should be
determined very precisely by a measurement at baseline around $\sim$10
m as demonstrated in \cite{mina-uchi}.  We thus do not discuss it
further by just assuming the setting with 10 location measurement
called the run IIB in \cite{mina-uchi}.  For the second stage, we
examine the setting with detectors placed at five locations $L = 350
\pm (0, 5, 10)$ m from the source, which covers a complete atmospheric
oscillation length.  The results of these two stages should be
combined as we will discuss bellow.

\begin{figure}[b]
\vglue -0.20cm
\begin{center}
\hglue  -0.5cm
\includegraphics[width=0.56\textwidth]{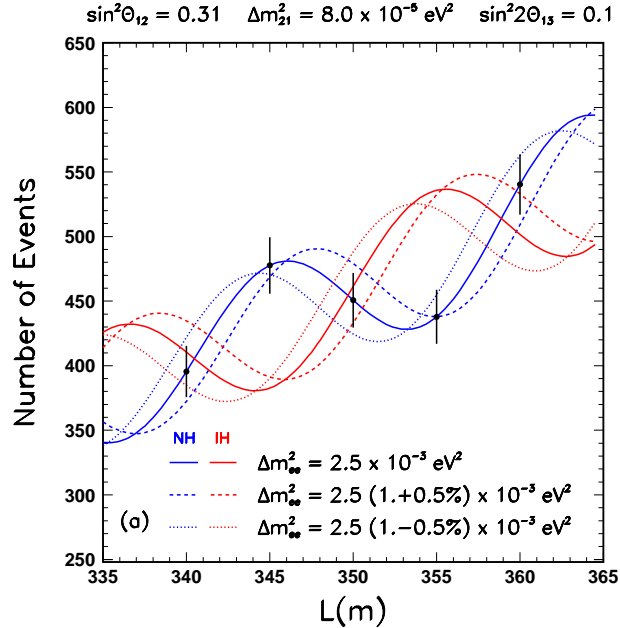}
\end{center}
\vglue -0.6cm
\caption{
Plotted are the expected number of events to be collected by detectors 
placed at the distances 340, 345, 350, 355 and 360 m from the source
for $\Delta m^2_{\text{ee}}=2.5 \times 10^{-3}$ eV$^2$ 
and $\sin^2 2\theta_{13} = 0.1$ for the normal hierarchy, 
indicated by solid circles with error bars. 
Here $P_\odot$ is included (solar parameters are fixed to 
their best fit values). 
Expected number of events in the absence of oscillation is assumed
to be 2000 at each detector position. 
The blue (light gray) solid curve which passes through the data points 
indicates a theoretical expectation 
assuming the normal hierarchy with 
$\Delta m^2_{\text{ee}}=2.5 \times 10^{-3}$ eV$^2$
and $\sin^2 2\theta_{13} = 0.1$ 
whereas the red (dark gray) curve off the data points is the one of 
inverted hierarchy with same values of $\Delta m^2_{\text{ee}}$
and $\theta_{13}$.  
Assuming that $\Delta m^2_{\text{ee}}$ is known to $\pm0.5$\% precision,
the thick-dotted (thin-dotted) lines above and below the solid lines 
are the expectations with $\Delta m^2_{\text{ee}}$ 0.5\% above (below) 
that value. 
}
\label{5point}
\end{figure}

In Fig.~\ref{5point} we present the expected number of events for
$\Delta m^2_{\text{ee}}=2.5 \times 10^{-3}$ eV$^2$ and $\sin^2
2\theta_{13} = 0.1$ at 5 locations mentioned above for the normal
hierarchy, for 2000 events at each detector location in the
absence of oscillation, which correspond to an exposure of about 20
MCi$\cdot$g$\cdot$day.\footnote{
In view of Fig.~\ref{5point}, one may think that 
measurement at least 3 different baseline distances would be 
enough to detect relative phases of the beat and the wiggle around it. 
Though it is true, it turned out that the five point measurement 
can achieve better accuracy than the three point measurement 
for an equal total number of events.
}
%
Throughout our analysis except for that in Sec.~\ref{sec:solar}, 
we fix the solar parameters to their best fit values as 
$\sin^2\theta_{12} = 0.31$ and $\Delta m^2_{21} = 8.0 \times 10^{-5}$ eV$^2$
and do not try to vary them. 
$\Delta m^2_{\text{ee}}$ is taken with a tentative expectations 
for its accuracy, 
$\Delta m^2_{\text{ee}}=2.5 \times 10^{-3}(1\pm 0.5$ \%) eV$^2$. 
The case of the normal and the inverted hierarchy are plotted by the 
blue (light gray) and the red (dark gray) curves, respectively. 
It appears that, if enough statistics is accumulated and 
$\Delta m^2_{\text{ee}}$ is separately measured accurately, the five point 
measurement, can discriminate between the normal and the inverted 
hierarchies.
The details of the statistical procedure is described in the next subsection.

\begin{figure}[thb]
\begin{center}
\hglue  -0.5cm
\includegraphics[width=0.7\textwidth]{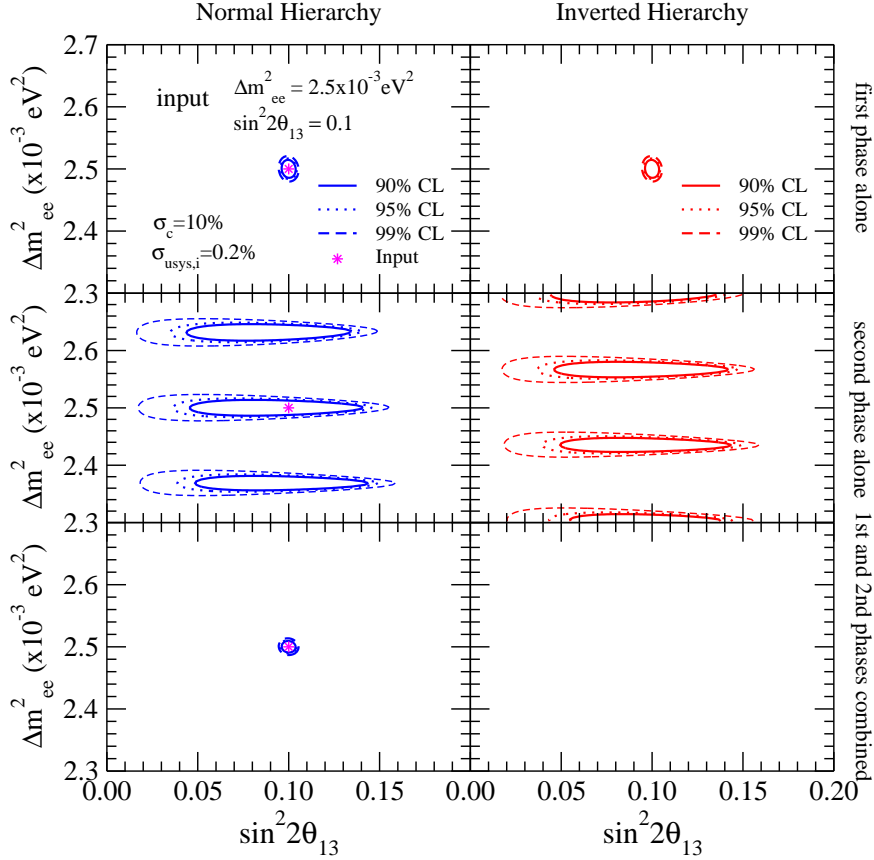}
\end{center}
\vglue -0.5cm
\caption{ The allowed regions in the $\sin^2 2\theta_{13} - \Delta
m^2_{\text{ee}}$ plane obtained by the first stage alone (upper
panels), the second stage alone (middle panels) and the combined
result of these two stages (lower panels) for the case where the input
parameters are $\sin^2 2\theta_{13} = 0.1$, $\Delta
m^2_{\text{ee}}=2.5 \times 10^{-3}$ eV$^2$ with the normal hierarchy,
based on the $\chi^2$ analysis described in 
Sec.~\ref{method}.  The solar parameters are fixed to their current
best fit values.  The first stage is assumed to be performed at 10
positions at around $L \sim 9$ m following the run IIB in
\cite{mina-uchi} whereas the second stage is assumed to be performed
at 5 locations at $L = 350 \pm (0, 5, 10)$ m from the source.  The
three closed curves (red, blue, green) from inner to outer denote the
ones obtained at 90\% CL, 95\% CL, and 99 \% CL.  }
\label{allowed-region}
\end{figure}

We show in three panels in Fig.~\ref{allowed-region} the allowed
regions in the $\sin^2 2\theta_{13} - \Delta m^2_{\text{ee}}$ space
obtained by the first stage alone (upper panels), by the second stage
alone (middle panels), and the one obtained when the first and the
second stages are combined (lower panels).  The input parameters are
taken as $\sin^2 2\theta_{13} = 0.1$, $\Delta m^2_{\text{ee}}=2.5
\times 10^{-3}$ eV$^2$, and the mass hierarchy is assumed to be the
normal one.
In the first-stage measurement to be done at around the distance 
$L_0/2 \simeq 10$ m from the source, 
$\sin^2 2\theta_{13}$ as well as $\Delta m^2_{\text{ee}}$
can be determined rather precisely. 
But, there is no sensitivity to the mass hierarchy because the baseline 
is too short, as is the case of reactor $\theta_{13}$ experiment.

The feature of the second-stage experiment presented in the 
middle panels of Fig.~\ref{allowed-region} requires some explanations. 
First of all, there exist many isolated island-like allowed regions. 
Suppose that $\theta_{13}$ is given from some other experiments. 
Then, due to the cosine term in Eq.~(\ref{eqn:Pdis}) the measurement 
will give a periodic solution
\begin{equation}
\Delta m'^2_{\text{ee}} \approx \Delta m^2_{\text{ee}} + \frac{4n \pi E}{L},
\ \ (n = \pm 1, \pm 2, ...) 
\end{equation}
for both of the hierarchies.  If we take approximation
$\phi_{\odot}=\pi/2$, there is a shift $2 \pi E / L$ between the
adjacent $\Delta m^2_{\text{ee}}$ solutions for the normal and the
inverted hierarchies.  Hence, the solutions of $\Delta
m^2_{\text{ee}}$ are alternating in the normal and the inverted
hierarchies with a constant shift $2 \pi E / L$ between the adjacent
solutions, the feature clearly visible in the middle panels of
Fig.~\ref{allowed-region}.  
The contours have finite widths, though prolonged, along the 
$\sin^2 2\theta_{13}$ direction because the second stage measurement
alone can determine $\theta_{13}$ but only with a limited accuracy.

It's clear that the second stage alone can not determine neither the
value of $\Delta m^2_{\text{ee}}$ nor the mass hierarchy.  However,
once the results of the first and the second stages are combined all
the fake solutions are eliminated, as demonstrated in the lower panels
of Fig.~\ref{allowed-region}.  It is evident that $\Delta
m^2_{\text{ee}}$ must be determined with high accuracy in the first
stage measurement.

\subsection{Analysis method and definition of $\chi^2$}
\label{method}

Now we give details of our analysis method. 
To calculate the sensitivity of the determination of the neutrino mass hierarchy 
by the first and second stages of the experiment combined,  
we compute 
\begin{equation}
\Delta \chi^2_{\text{min}}(\sin^2 2\theta_{13}) = \chi^2_{\text{min}}(\rm wrong \; hierarchy)-\chi^2_{\text min}(\rm true \; hierarchy)\, ,
\end{equation}
where $\chi^2_{\text {min}}(\rm true/wrong \; hierarchy)$ is the minimum of  
\begin{equation}
\chi^2(\rm true/wrong \; hierarchy)= \chi^2_1 + \chi^2_2 (\rm true/wrong \; hierarchy) \, ,
\end{equation}
with the first stage  $\chi^2$ given by
\begin{equation}
\chi^2_{\text{1st}}(\Delta m^2_{\text{ee}},\, \sin^2 2\theta_{13}) 
=\sum_{i,j=1}^{10} 
\left[
\frac{ N_i^{\text{1-obs}}-N_{i}^{\text{1-theo}}}{N_i^{\text{1-theo}}}\right]\, 
(V^{-1}_{\text{1}})_{ij}\, 
\left[
\frac{N_j^{\text{1-obs}}-N_{j}^{\text{1-theo}}}{N_j^{\text{1-theo}}}\right]
\, ,
\end{equation}
and the second stage $\chi^2$ given by
\begin{equation}
\chi^2_{\text{2nd}}(\Delta m^2_{\text{ee}},\, 
\sin^2 2\theta_{13}, \text{sign}(\Delta m^2_{31})) 
=\sum_{i,j=1}^{5} 
\left[
\frac{N_i^{\text{2-obs}}-N_{i}^{\text{2-theo}}}
{N_i^{\text{2-theo}}}\right]\, 
(V^{-1}_{\text{2-obs}})_{ij}\, 
\left[
\frac{(N_j^{\text{2}}-N_{j}^{\text{2-theo}})}{N_j^{\text{2-theo}}}\right]. 
\end{equation}
$N_i^{\text{1,2-obs}}$ is the number of events at baseline $L_i$ for a given 
input hierarchy and values of the mixing parameters (observed events), while
 $N_i^{\text{1,2-theo}}$ is the theoretically expected number of events 
at baseline $L_i$ for a given set of oscillation parameters.
We use, as in Ref.~\cite{mina-uchi}, the correlation matrices 
\begin{equation}
(V^{-1}_{\text{1,2}})_{ij}= \frac{\delta_{ij}}{\sigma^2_{ui}} - 
\frac{1}{\sigma^2_{ui}\sigma^2_{uj}}
\frac{\sigma^2_c}{[1+(\sum_{k} \frac{1}{\sigma^2_{uk}})\sigma^2_c]},
\end{equation}
where 
$\sigma^2_{ui} = \sigma^2_{\text{usys}} + 1 / N_{i}^{\text{1,2-th}}$. 

The estimation of the uncorrelated systematic uncertainty is the key 
issue in our sensitivity analysis. 
In \cite{mina-uchi} two cases of the error, 
$\sigma_{\text{usys}}=0.2$ \% and 1\%, 
are examined. The former corresponds to the case of movable detector, 
assuming that a direct extraction or real-time detection of produced 
$^3$H atom is possible. 
The latter is a tentative estimate for the case of multiple identical 
detectors.  
We use the correlated systematic uncertainty $\sigma_c=10$ \%, 
though the results are insensitive to $\sigma_c$. 
We adopt the same numbers as these for the uncertainties in our analysis 
for both the first and the second stages of the experiment.

In our $\chi^2$ analysis, for simplicity, the solar mixing parameters 
are fixed to the current best fit values, ignoring their uncertainties. 
%
Therefore, the fitting parameters are: 
$\Delta m^2_{\text{ee}}$, $\sin^2 2\theta_{13}$ and the sign of 
$\Delta m^2_{31}$. 
Note, however, that $\chi^2_{\text{1st}}$ for the first stage has practically 
no dependence on solar parameters as well as the mass hierarchy 
due to too short baseline. 
The current uncertainty on the values of the solar parameter 
are not so relevant  for the determination for the hierarchy 
because $\phi_\odot$ does not depend strongly on 
them (see Fig.~\ref{fig:phi}b) as pointed out previously.

\subsection{Sensitivity to mass hierarchy determination}
\label{subsec:res}

\begin{figure}[bhtp]
\vglue -0.2cm
\includegraphics[width=0.52\textwidth]{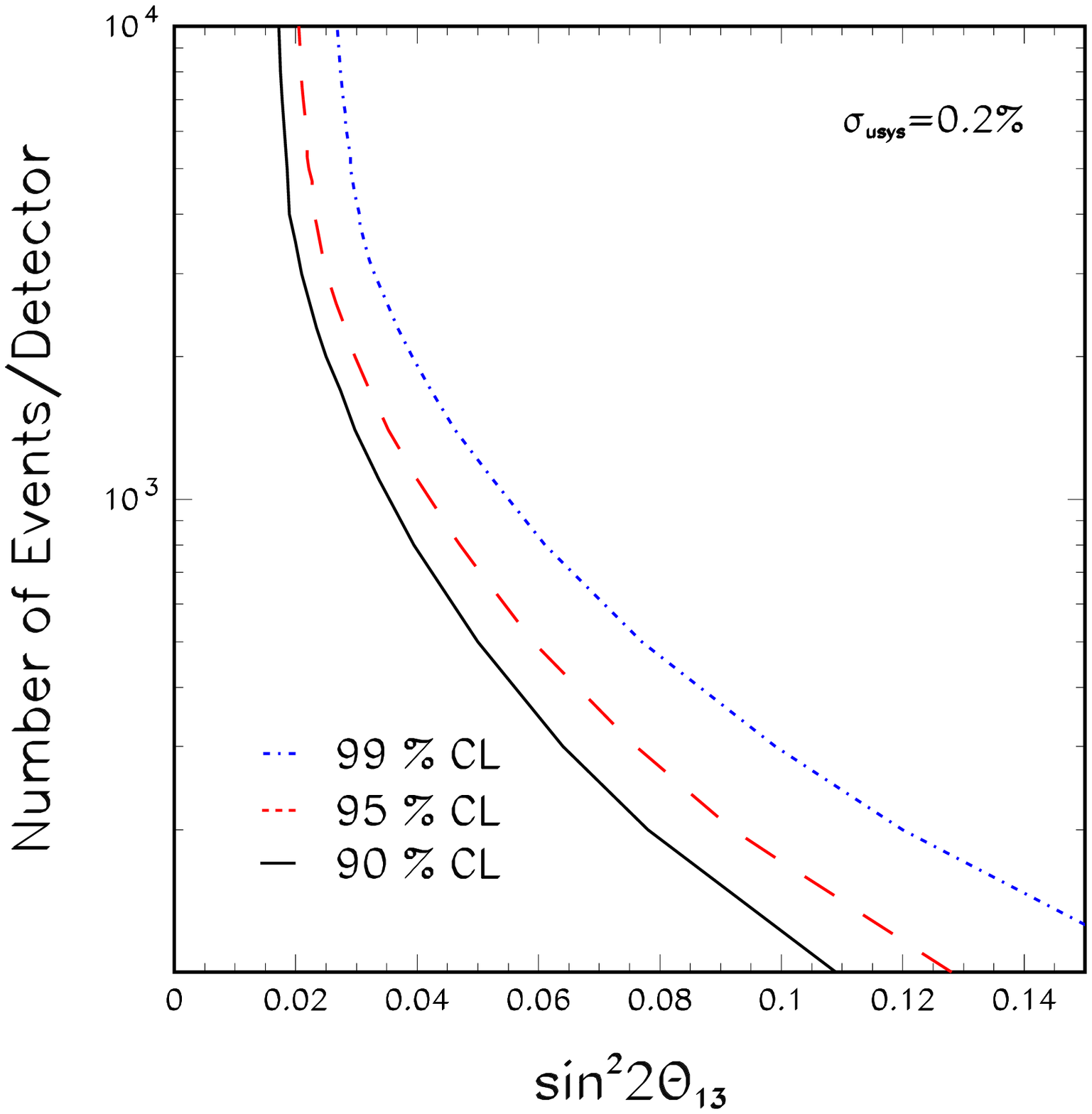}
\hglue  -0.8cm
\includegraphics[width=0.52\textwidth]{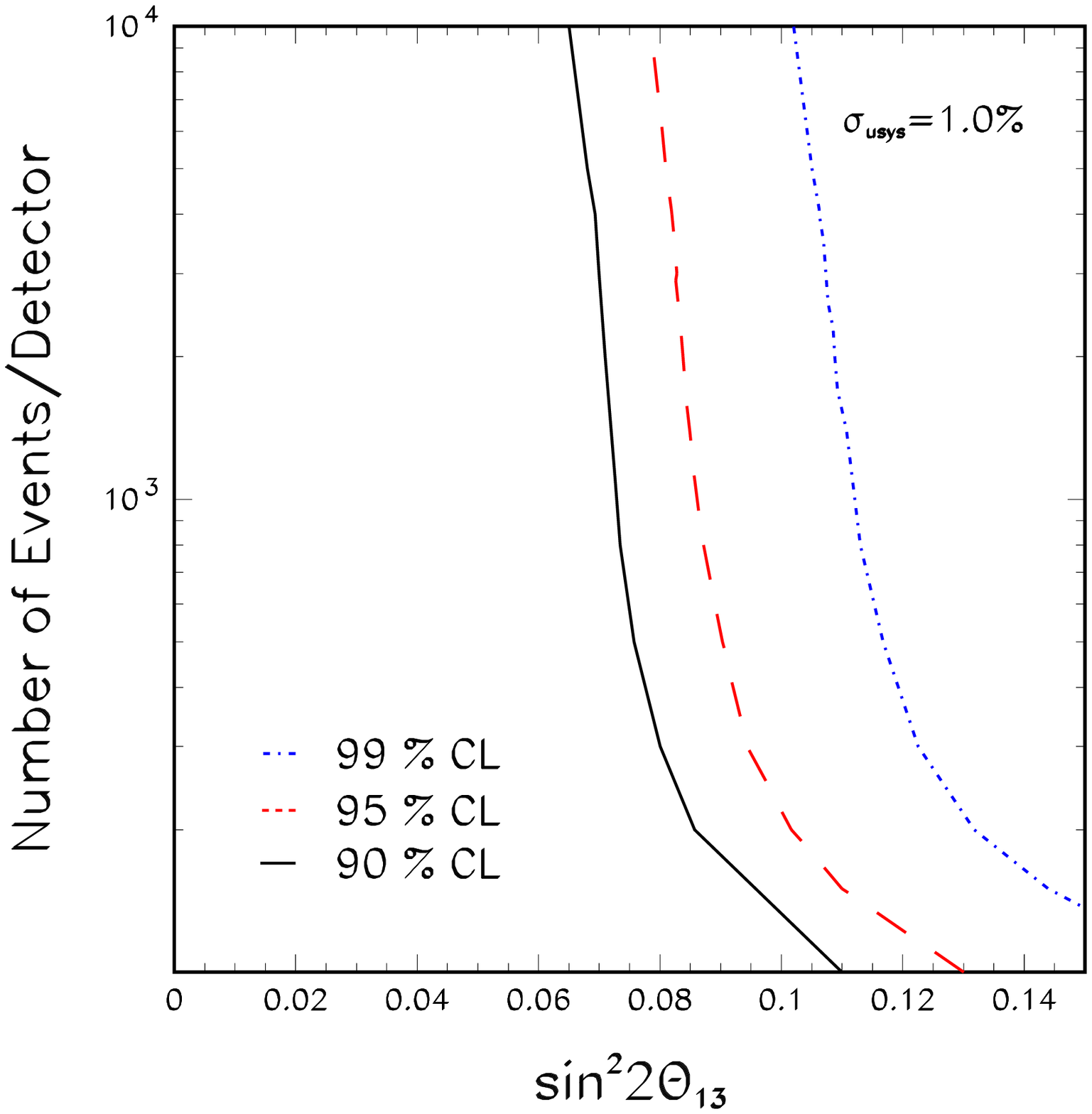}
\vglue -0.6cm
\caption{The region of sensitivity to resolving the mass hierarchy in 
$\sin^2 2\theta_{13} -$event number (per detector) space. 
The black solid, the red dashed, and the blue dotted curves 
denote the region boundary at 90\%, 95\%, and 99\% CL, respectively. 
The left and right panels are for the cases of 
uncorrelated systematic uncertainties of 0.2\% and 1\%, respectively. 
}
\label{fig:sensitivity}
\end{figure}

Now, we present our estimation of the sensitivity of the two-stage
M\"ossbauer $\bar \nu_e$ experiment to the determination of the
neutrino mass hierarchy.
In Fig.~\ref{fig:sensitivity} we present the region of sensitivity to
resolving the mass hierarchy at 90\%, 95\%, and 99\% CL, denoted by
the black solid, the red dashed, and the blue dotted curves,
respectively.  The abscissa is $\sin^2 2\theta_{13}$ and the ordinate
is the number of events per detector in long baseline experiment.
Here, 90\%, 95\%, and 99\% CL correspond to $\Delta
\chi^2_{\text{min}}= \chi^2_{\text{min}}(\rm wrong \; hierarchy)-
\chi^2_{\text min}(\rm true \; hierarchy)$ = 2.71, 3.84 and 6.63,
respectively.
Left and right panels are for the cases where the uncorrelated systematic
uncertainty is 0.2\% and 1\%, respectively.

Assuming $\sigma_{\text{usys}}=0.2$ \%, we can see that the hierarchy 
can be determined at 90 (99) \% CL for $\sin^2 2\theta_{13}$ 
down to 0.035 (0.06) for 1000 events and 
down to 0.025 (0.04) for 2000 events.
For a larger error $\sigma_{\text{usys}}=1$\%, the critical value of 
$\sin^2 2\theta_{13}$ will be $\sim$ 0.08 (0.11) at 90 (99) \% CL 
approximately independent of the number of events larger than 
$\sim$ 1000. 
This is caused by the fact that 
the precision on the determination of $\Delta m^2_{\text{ee}}$
at the first stage is highly dependent on this systematic 
uncertainty~\cite{mina-uchi}.

\section{Determination of the Solar Parameters}
\label{sec:solar}

In the second stage of the M\"ossbauer experiment, 
the detectors are placed just after the first solar oscillation maximum
in order to determine the mass hierarchy, as we discussed in 
the previous section. 
In a possible third stage of the experiment, one could envisage  
moving the detectors somewhat closer to the source in order to 
cover the region around this maximum. 
It will allow us a precise determination of the solar-scale 
oscillation parameters, 
$\Delta m^2_{21}$ and $\theta_{12}$.

In order to optimize the determination of these parameters, 
we assume that the measurements will be performed at the following 
10 different detector locations, 
\begin{subequations}
\begin{eqnarray} 
L_n& =& [200+50\,(n-1)] \; {\rm m},  
\label{location-solar1} \\
L'_n& =& L_n-L_0/2,\quad  \quad n=1,2,..,5.  
\label{location-solar2}
\end{eqnarray} 
\end{subequations}
First, in order to observe the oscillation driven by 
the solar $\Delta m_{21}^2$, we consider the configuration
of 5 detector locations ($L_n$) separated by 50 m, ranging from 
200 m to 400 m as in Eq. (\ref{location-solar1}).
In this way, we can cover the whole range of solar-scale oscillation 
before and after the dip due to the oscillation maximum 
(see Fig.~\ref{Rio-SP}). 
Second, in order to minimize the unwanted effect due to the rapid 
oscillations driven by the atmospheric $\Delta m_{\text{ee}}^2$, 
we have to place, for each location $L_n$, another 
detector (or move detector to another location) at 
$L'_n$ separated from $L_n$ by half the atmospheric oscillation 
length, $L_0/2$. See Eq. (\ref{location-solar2}). 
The setting of five pairs of detector locations 
works because the average of the oscillation probabilities 
corresponding to these two positions $L_n$ and $L'_n$ is, 
at first approximation, 
independent of $\theta_{13}$ and $\Delta m_{\text{ee}}^2$ 
and thus is free from 
any ambiguity associated with the atmospheric oscillations.

In this calculation we assume that the exposure is such 
that the same number of events (1000) in the absence of oscillation 
are collected at each detector location. 
We take the pessimistic value of the systematic uncertainty 
$\sigma_{\rm usys}=$ 1\% instead of 0.2 \% because the results 
do not differ much; 
The sensitivity is still dominated by the statistics in this regime. 
As in the previous sections, the input values of the other oscillation parameters 
are fixed as 
$\Delta m^2_{\text{ee}}=2.5\times 10^{-3}$ eV$^2$,
$\sin^2\theta_{12} = 0.31$ and $\Delta m^2_{21} = 8.0 \times 10^{-5}$ eV$^2$.  
%

\begin{figure}[bhtp]
\begin{center}
\vglue  -0.3cm
\includegraphics[width=0.6\textwidth]{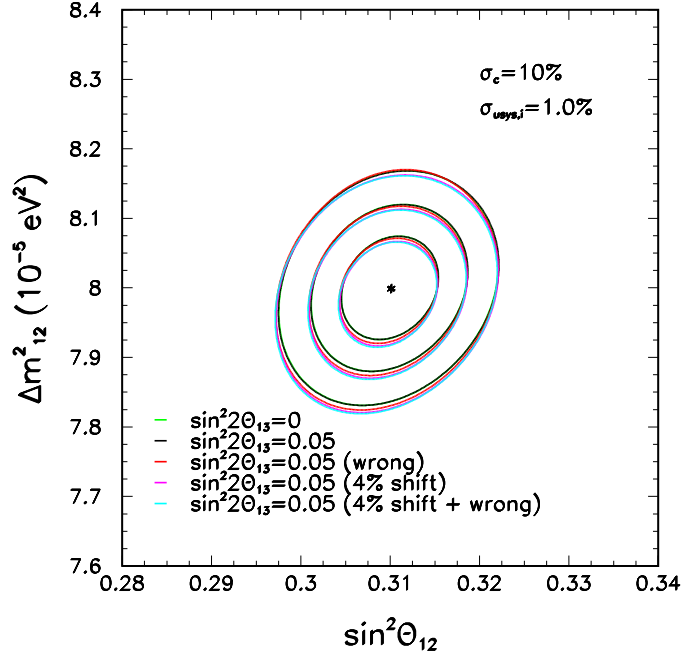}
\end{center}
\vglue -0.4cm
\caption{Expected allowed regions in the plane
  $\sin^2\theta_{12}-\Delta m^2_{21}$ for 1000 events taken at each
  detector position at 68.27\% (inner curves), 95.0\% (middle curves)
  and 99.73\% (outer curves) CL for 2 DOF.  We assumed the measurement
  will be done at 10 different detector locations, $L_n = [200 +
  50\; (n-1)]$~m and $L'_n = L_n-L_0/2$ with $n=1,2,...,5$.  The input values
  used are: $\sin^2\theta_{12}=0.31$, $\Delta m^2_{21} = 8.0 \times
  10^{-5}$ eV$^2$, $\Delta m^2_{\text{ee}}=2.5\times 10^{-3}$ eV$^2$.
  The green and black curves are for the input values of $\sin^2
  2\theta_{13}=0$ and $0.05$, respectively.  The other curves are also 
  for $\sin^2 2\theta_{13}=0.05$ but fitted with: the wrong
  hierarchy (red), $\Delta m^2_{\text{ee}}$ shifted by 4\%
  (magenta) and the wrong hierarchy and $\Delta
  m^2_{\text{ee}}$ shifted by 4\% (cyan).  Here we have supposed 
  $\sigma_{\text{usys}}=$ 1\%.
}
\label{fig:solar}
\end{figure}

In Fig.~\ref{fig:solar} we show how well $\Delta m^2_{21}$ and 
$\sin^2\theta_{12}$ can be determined by this experiment. 
The obtained sensitivity contours with $\sin^2\theta_{13}=0$ 
are presented by the green lines (which practically 
overlapped with the black ones) in Fig.~\ref{fig:solar}. 
From this result, we conclude that the precision one can 
achieve for $\sin^2\theta_{12}$ is about 
1.6 (4.0) \% at 68.27 (99.73) \% CL for 2 DOF
whereas that for $\Delta m^2_{21}$ is about 
0.9 (2.2) \% at 68.27 (99.73) \% CL for 2 DOF. 
Comparing this result with the sensitivity which can be achieved by
the dedicated reactor experiment for the similar number of events
studied in Ref.~\cite{sado+others}, the precision we obtained for
$\sin^2\theta_{12}$ and $\Delta m^2_{21}$ in this work are comparable
or better, in particular, for $\Delta m^2_{21}$ (see e.g. Fig. 5 of
the first article in ~\cite{sado+others}).

\begin{figure}[bhtp]
\begin{center}
\vglue  -0.5cm
\includegraphics[width=0.66\textwidth]{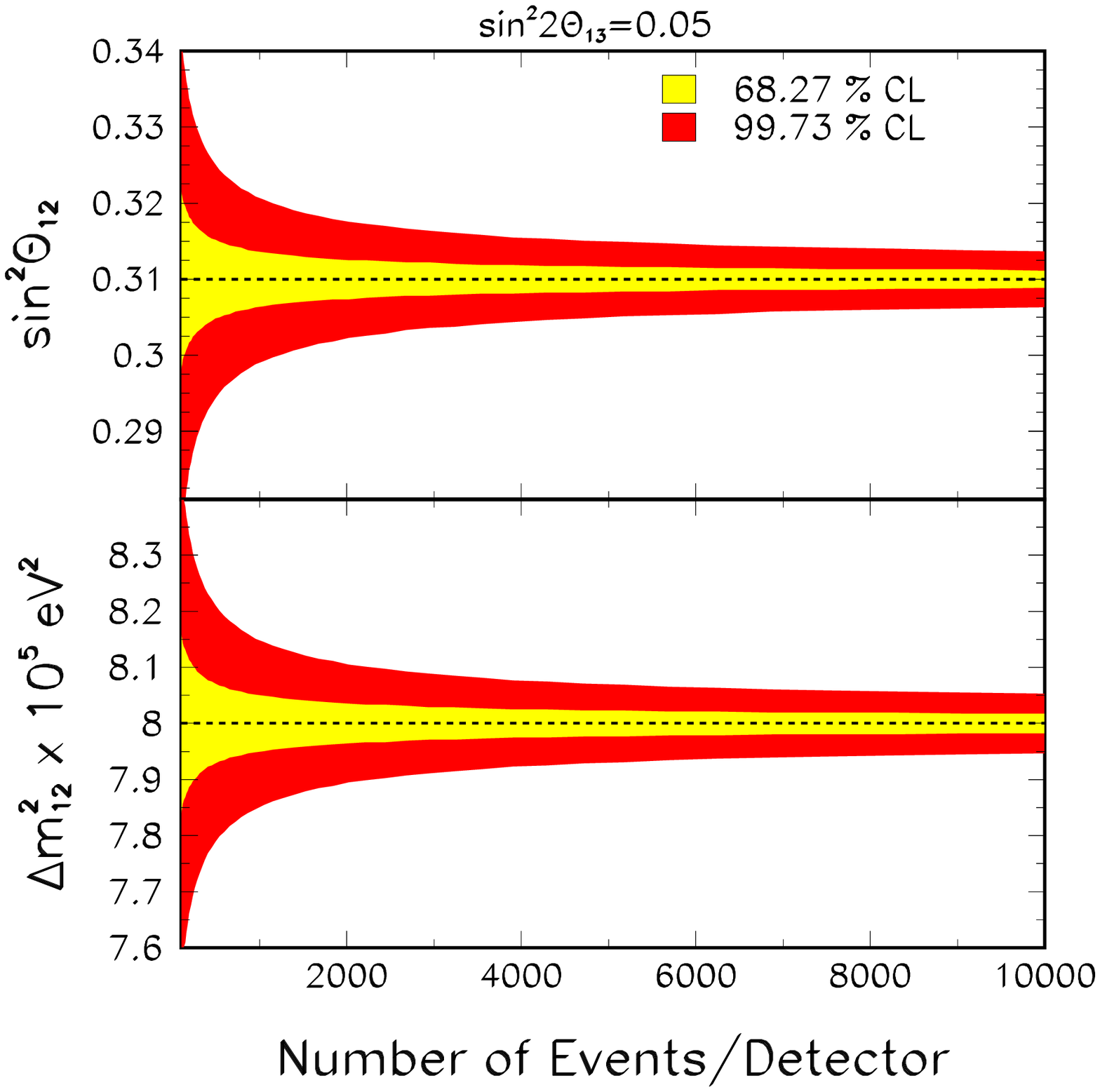}
\end{center}
\vglue -0.8cm
\caption{Determination of $\sin^2\theta_{12}$ (upper panel) and
$\Delta m^2_{21}$ (lower panel) as a function of detector 
exposure at 68.27\% and 99.73\% CL for 1 DOF.
Here $\sigma_{\rm usys}=1\%$. The dashed lines show the 
input values. }
\label{fig:dms12}
\end{figure}

In the above argument, we have assumed that 
$L_0$ (or $\Delta m_{\text{ee}}^2$) 
can be determined very precisely in the first stage of the experiment allowing 
us to place the detector pair separated exactly by $L_0/2$. 
In reality, however, the determination of $L_0$ will always carry 
some uncertainties. 
Moreover, if $\sin^2 2\theta_{13}$ is too small or 
the uncorrelated systematic uncertainty ($\sigma_{\text{usys}}$) 
is not small enough, 
the neutrino mass hierarchy will not be determined in the 
second stage. 
Then, the question is; What is the additional ambiguities that 
occur in determination of the solar oscillation 
parameters in the presence of these uncertainties? 
Fortunately, we can show that our choice of detector locations defined in 
Eqs.~(\ref{location-solar1}) and (\ref{location-solar2}) 
guarantees that these uncertainties will not deteriorate 
the precision in the determination of the solar parameters.

In order to check the stability of the results against 
the uncertainties mentioned above, we 
performed a fit by assuming the incorrect mass hierarchy 
and/or slightly incorrect value of $\Delta m^2_{\text{ee}}$ 
by shifting its value by 4\%. 
The results are shown by the blue, the magenta and the light blue curves 
in Fig.~\ref{fig:solar} where we can see that for each CL, 
all the curves are very close to each other, 
indicating that our results are quite 
stable against these uncertainties.

We also examine the stability of the sensitivities to possible nonzero
values of $\theta_{13}$.  In Fig.~\ref{fig:solar} we show the
sensitivity contours for the input values of $\sin^2
2\theta_{13}=0.05$ (black curve) in comparison to that with $\sin^2
2\theta_{13}=0.0$ (green curve).
We note that these 2 curves are very close to each other implying that
the effect of nonzero $\theta_{13}$ is quite small.  We emphasize that
this feature is in a marked contrast with any of the alternative
methods for measuring $\theta_{12}$, including the reactor
\cite{sado+others} and the solar neutrino methods \cite{nakahata}.  In
these methods there is no way, unless determined by some other
experiments, to control the errors which comes from nonzero
$\theta_{13}$ \cite{concha-pena}, whereas in our present method,
there is a built-in mechanism which eliminates the uncertainties that
come from the atmospheric oscillation parameters, in particular from
$\theta_{13}$.

In Fig.~\ref{fig:dms12} we show how the precision for 1 DOF of each of
the solar parameters improves with statistics, that is, as a function
of the number of events that would be collected at each detector in
the absence of oscillation.  As in the previous plot, we have assumed
that $\sigma_{\rm usys}=$ 1\%.
We observe that until $\sim 1000$ events, the precision 
improves rapidly as the number of events increase and then after,
the rapidness of the improvement become smaller.
The ultimate precision which can be achieved in the limit of very large
statistics with $\sim 10000$ events (per detector) for $\sigma_{\rm usys}=$ 1\%
would be $\sim$ 0.4 (1.2) \% for 68.27\% (99.73\%) CL 
for $\sin^2 \theta_{12}$ and 
$\sim$ 0.2 (0.7) \% for 68.27\% (99.73\%) CL for $\Delta m^2_{21}$. 
Notice that measurement of $\Delta m^2$ to the accuracy of 
sub \% level is usually not possible because the absolute energy scale 
of leptons is hard to be determined at such accuracy, and it acts as a 
limiting factor for reducing the $\Delta m^2$ error \cite{MNPZ}. 
In our case, monochromaticity of the $\bar{\nu}_{e}$ beam with known 
energy removes the necessity for such calibration for determining 
absolute energy scale.

\section{Concluding Remarks}
\label{sec:conc}

In this paper, we have shown that the neutrino mass hierarchy can be
determined only by {\it a single experiment} based on resonant $\bar\nu_e$
absorption reaction enhanced by the M\"ossbauer effect.  In its first
 stage, $\Delta m^2_{\text{ee}}$ must be accurately determined by $L
\sim$10 m measurements.  In the second stage of the experiment, the
relative phase of the atmospheric oscillation (wiggles) between the
normal vs. inverted mass hierarchies, phase advancement or
retardation must be discriminated by measurements at $L \sim$350 m.
It can be executed thanks to the ultra monochromatic nature of 
neutrino beam which would allow us to detect relative phase of 
atmospheric oscillation even after 20 oscillations. 
We stress that the monochromaticity is special to the resonant 
$\bar\nu_e$ absorption reaction, 
most probably the unique experimental way to realize this principle. 

We concluded that assuming $\sigma_{\text{usys}}=0.2$ \%, the
hierarchy can be determined at 90 (99) \% CL for $\sin^2 2\theta_{13}$
down to 0.035 (0.06) for 1000 events collected at five different positions
and down to 0.025 (0.04) for 2000 events. 
For a larger error $\sigma_{\text{usys}}=1$\%, the critical value of
$\sin^2 2\theta_{13}$ will be $\sim$ 0.08 (0.11) at 90 (99) \% CL
approximately independent of the number of events for more than $\sim$
1000 events.

We have also shown that in a third stage of the experiment, by
positioning detectors in different locations, it is possible to
determine the solar-scale parameters too. 
The precision one can achieve with 1000 events in each detector 
for $\sin^2\theta_{21}$  is about 1.6 (4.0) \% at 68.27 (99.73) \% CL 
whereas that for 
$\Delta m^2_{21}$ is about 0.9 (2.2) \% at 68.27 (99.73) \% CL, all for 2 DOF.
The accuracies improve as data taking proceeds and the errors 
would reduce by a factor of $\simeq3$ when an order of magnitude 
larger number of events is taken.

It is in fact remarkable that the experimental set up considered 
in this paper allows us to make a precise measurement of 
all the mixing parameters appear in the oscillation probability for 
$\nu_e \to \nu_e$ channel, 
$\theta_{12}$, $\theta_{13}$, $\Delta m^2_{21}$, 
$\Delta m^2_{\text{ee}}$ and the neutrino mass hierarchy 
only by a single experiment, independent of any other experiments. 
The only parameters inaccessible by the method are 
$\theta_{23}$ and the CP phase $\delta$, 
but precision measurement of the other mixing parameters
will certainly help in determining these parameters by other
experiments.

Some remarks are in order: 

\begin{itemize}

\item
As noted in Sec.~\ref{subsec:theo}, our method of determining the mass
hierarchy can be confused by the CPT violation of the type with
$\theta_{12}$ in different octants in the neutrino and the
antineutrino sectors.  It in turn implies that once the mass hierarchy
is determined by some other means, the measurement we propose in this
paper can be regarded as testing that particular type of CPT violation.

\item
Due to the capability of resolving atmospheric wiggles 
even after 20 atmospheric oscillations away from the source, 
this experiment could
be sensitive to other dynamics~\cite{various} which differ from the
standard three flavor mass induced oscillation one.  In particular,
one can perhaps probe mechanisms such as: neutrino decay~\cite{decay},
decoherence~\cite{decoh}, mixing with sterile
neutrinos~\cite{sterile}, violation of Lorentz invariance or the 
equivalence principle~\cite{VEPVLI}, long-range leptonic forces~\cite{LR},
extra-dimensions~\cite{extradim}.

\end{itemize}

Finally, we emphasize that realizing the proposed M\"ossbauer 
$\bar\nu_e$ experiment is crucial in order to make all the discussions 
in this paper meaningful. 
The principle of the experiment is well formulated by having 
the decay producing source $\bar\nu_e$ and the absorption reaction 
in the target as T-conjugate channels with each other, allowing 
mutual cancellation of large part of the atomic shift 
\cite{Raghavan,raghavan_now2006}. 
Nonetheless, it still remains unclear to what extent the $^3$He atom 
in the target experiences exactly the same environment as the $^3$H atom 
in the source \cite{potzel_snow2006}. 
This point has to be (and will be) investigated in a systematic way. 
If the experiment is indeed realizable it will certainly open a
new window for neutrino experiments which can explore the 
neutrino properties in otherwise inaccessible manner. 


\begin{acknowledgments}
  One of the authors (H.M.) thanks the Abdus Salam International
  Center for Theoretical Physics, where part of this work was done,
  for the hospitality.  Two of us (H.N. and R.Z.F) are grateful for
  the hospitality of the Theory Group of the Fermi National
  Accelerator Laboratory during the summer of 2006.
  This work was supported in part by the Grant-in-Aid for Scientific Research, 
  Nos. 16340078 and 19340062, Japan Society for the Promotion of Science,
  by Funda\c{c}\~ao de Amparo \`a Pesquisa do Estado de S\~ao Paulo
  (FAPESP), Funda\c{c}\~ao de Amparo \`a Pesquisa do Estado de Rio de
  Janeiro (FAPERJ) and Conselho Nacional de Ci\^encia e Tecnologia
  (CNPq). Fermilab is operated under DOE Contract No.
  DE-AC02-76CH03000.
\end{acknowledgments}

\appendix


\section{The Effective Atmospheric $\Delta m^2$ }
\label{appendixA}
\begin{figure}[bhtp]
\begin{center}
\includegraphics[width=0.55\textwidth]{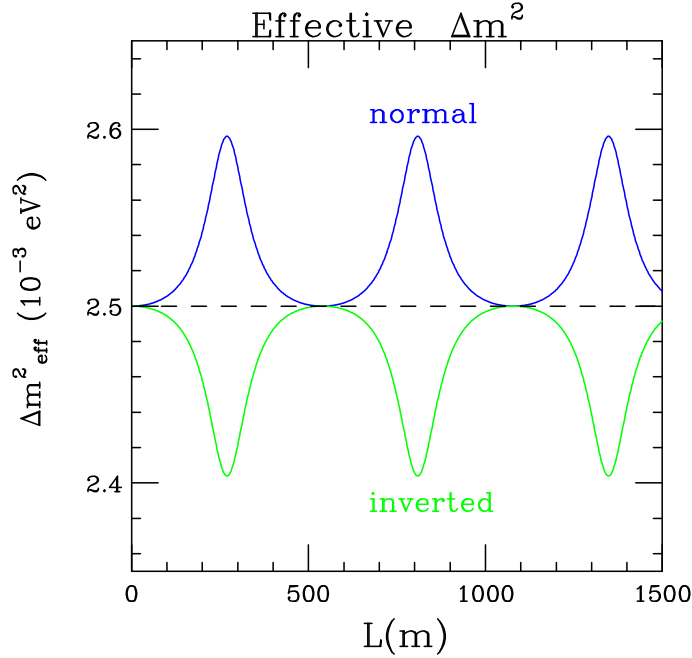}
\end{center}
\caption{$\Delta m^2_{\text{eff}}$ as function of baseline at 18.6 keV
neutrino energy for both the normal and inverted hierarchies.  The
dashed line is $ \Delta m^2_{\text{ee}}$ which has been assumed to be
the same for the normal and inverted hierarchies for this plot. In
general, unless $ \Delta m^2_{\text{ee}}$ is determined precisely it
need not take exactly the same value for both hierarchies.  The
maximum difference between $\Delta m^2_{\text{eff}}$ and $ \Delta
m^2_{\text{ee}}$ occurs at the solar oscillation maxima, i.e. when
$\Delta_{21} = (2n-1)\pi/2$.  }
\label{fig:dmsq}
\end{figure}

The advancement or retardation of the phase, $\phi_\odot$, could be
interpreted as a change in the effective atmospheric $\Delta m^2$.
Such an effective atmospheric $\Delta m^2$, at $L/E$, is naturally
defined by
\begin{eqnarray}
\Delta m^2_{\text{eff}}(L/E) & \equiv & 
\frac{d~(2\Delta_{\text{ee}} \pm \phi_\odot)}{d~(L/2E)}  
=  \Delta m^2_{\text{ee}} \pm \frac{1}{2}
~\Delta m^2_{21}\cos 2 \theta_{12} 
~\frac{ \sin^2 2 \theta_{12} \sin^2 \Delta_{21} }
{1-  \sin^2 2 \theta_{12} \sin^2 \Delta_{21} } 
\label{eqn:dmsq_eff} \\[0.5cm]
& = & \left(\frac{c^2_{12} -\frac{1}{2}\sin^2 2 \theta_{12}\sin^2 \Delta_{21}}
 {1-  \sin^2 2 \theta_{12} \sin^2 \Delta_{21} }\right) \vert \Delta m^2_{31} 
\vert 
+ \left(\frac{s^2_{12} -\frac{1}{2}\sin^2 2 \theta_{12}\sin^2 \Delta_{21}}
 {1-  \sin^2 2 \theta_{12} \sin^2 \Delta_{21} } \right)\vert \Delta m^2_{32} \vert.
 \nonumber
\end{eqnarray}
Where the positive (negative) sign, in Eq.~(\ref{eqn:dmsq_eff}),  
is for the normal (inverted) hierarchy.
For $\Delta_{21} \ll 1$ we recover $\Delta m^2_{\text{ee}}$, since
this is in essence its definition, see Ref.~\cite{NPZ} for details.
Thus Eq.~(\ref{eqn:dmsq_eff}) is the generalization of 
$\Delta m^2_{\text{ee}}$ to arbitrary $L/E$.
In Fig. \ref{fig:dmsq} we show the variation of 
$\Delta m^2_{\text{eff}}$ as a function of baseline for both
the normal and inverted hierarchies.
The extrema of the difference between 
$\Delta m^2_{\text{eff}}(L/E)$ and $ \Delta m^2_{\text{ee}}$  are given by
\begin{eqnarray}
\hspace*{-0.5cm}
\frac{1}{2}
~\Delta m^2_{21}\cos 2 \theta_{12} 
~\frac{\sin^2 2 \theta_{12} \sin^2 \Delta_{21} }
{1-  \sin^2 2 \theta_{12} \sin^2 \Delta_{21} }
 & = & \left\{ \begin{array}{ll}
0 &{\rm at } ~ \Delta_{21}= n\pi \\
\frac{1}{2} ~\frac{\sin^2 2 \theta_{12}}{  \cos 2 \theta_{12}}
~\Delta m^2_{21}  &{\rm at } ~\Delta_{21}= (2n-1)\pi/2. 
\end{array} \right.
\end{eqnarray}
As the oscillations go through the solar maxima, the fractional
increase (decrease) in the effective atmospheric $\Delta m^2$ is
\begin{eqnarray}
\frac{1}{2} ~\frac{\sin^2 2 \theta_{12}}{  \cos 2 \theta_{12}}
~\frac{\Delta m^2_{21}}{\Delta m^2_{\text{ee}}}  \approx 4\%\, ,
\end{eqnarray}
for the normal (inverted) hierarchy.  
It is the accumulated effect of this
small difference in the effective atmospheric $\Delta m^2$ for the
normal and inverted hierarchies that leads to the observable phase
difference shown in Fig.~\ref{Rio-SP}. Since 
\begin{eqnarray}
 \frac{1}{2}\int_0^{L/E} d(L/E) ~\Delta
m^2_{\text{eff}}(L/E) =(2\Delta_{\text{ee}}\pm\phi_\odot)\vert_{L/E},
\end{eqnarray}
as it must.

\section{$\nu_\mu$ Disappearance}
\label{appendixB}

For $\nu_\mu$ disappearance in vacuum, most of the results of this paper can be applied,
using the following translation
\begin{eqnarray}
\sin^2 \theta_{12} \Rightarrow \frac{\vert U_{\mu 2}\vert^2}{\vert U_{\mu 1}\vert^2+\vert U_{\mu 2}\vert^2}
\quad  & {\rm and} & \quad \sin^2 \theta_{13} \Rightarrow \vert U_{\mu 3} \vert^2.
\end{eqnarray}
With this translation the atmospheric $\Delta m^2_{\mu\mu}$ and phase, $\phi^\mu_\odot$, are given by
\begin{eqnarray}
\Delta m^2_{\mu \mu} & \equiv &  \frac{\vert U_{\mu 1}\vert^2 \vert \Delta m^2_{31}  \vert + \vert U_{\mu 2}\vert^2   \vert \Delta m^2_{32}  \vert  }{\vert U_{\mu 1}\vert^2  + \vert U_{\mu 1}\vert^2}
\label{eqn:dmsqmumu}\\
{\rm and} \quad \phi^\mu_\odot  &= & \arctan \left(\frac{\vert U_{\mu 1}\vert^2 - \vert U_{\mu 2}\vert^2}{\vert U_{\mu 1}\vert^2 +\vert U_{\mu 2}\vert^2 } \tan \Delta_{21}\right) - \frac{\vert U_{\mu 1}\vert^2 - \vert U_{\mu 2}\vert^2}{\vert U_{\mu 1}\vert^2 +\vert U_{\mu 2}\vert^2 }\Delta_{21} .
\end{eqnarray}
Then
\begin{eqnarray}
 & &  P(\nu_\mu \rightarrow \nu_\mu) = 1  - 4\vert U_{\mu 2}\vert^2 \vert U_{\mu1}\vert^2 \sin^2 \Delta_{21}
\nonumber \\  & & \hspace{-1cm} 
-2\vert U_{\mu 3}\vert^2(1- \vert U_{\mu 3}\vert^2)
\left[ 1-\sqrt{ 1-\frac{4\vert U_{\mu 1}\vert^2 \vert U_{\mu 2}\vert^2}
{(\vert U_{\mu 1}\vert^2 +\vert U_{\mu 2}\vert^2 )^2}\sin^2 \Delta_{21}}
~\cos (2 \Delta_{\mu \mu} \pm \phi^\mu_\odot) \right] .
\label{eqn:Pdismu}
\end{eqnarray}

Since numerically
\begin{eqnarray}
\frac{\vert U_{\mu 1}\vert^2 - \vert U_{\mu 2}\vert^2}{\vert U_{\mu 1}\vert^2 +\vert U_{\mu 2}\vert^2 } \approx -\cos 2 \theta_{12}
\end{eqnarray}
the phase for $\nu_\mu$ disappearance is approximately minus the phase for
$\nu_e$ disappearance, i.e.
\begin{eqnarray}
\phi^\mu_\odot \approx - \phi^e_\odot.
\end{eqnarray}
Therefore the normal hierarchy is retarded while the inverted hierarchy is advanced. This is 
the opposite sign to what happens for $\nu_e$ disappearance but the rate of change is approximately
the same.

The effective atmospheric $\Delta m^2$ for $\nu_\mu$ disappearance at any $L/E$ is given by
\begin{eqnarray}
\Delta m^2_{\text{eff}}(\mu \mu) & = & 
\left[ \frac{\vert U_{\mu1} 
\vert^2(\vert U_{\mu1} \vert^2+\vert U_{\mu2} \vert^2)
-2\vert U_{\mu1} \vert^2\vert U_{\mu2} \vert^2 \sin^2 \Delta_{21}}
{(\vert U_{\mu1} \vert^2+\vert U_{\mu2} \vert^2)^2-4\vert U_{\mu1} \vert^2\vert U_{\mu2} \vert^2 \sin^2 \Delta_{21}} \right]\vert \Delta m^2_{31} \vert 
\nonumber \\[0.1cm]  
& &  \quad + \left[ \frac{\vert U_{\mu2} \vert^2(\vert U_{\mu1} \vert^2
+\vert U_{\mu2} \vert^2)
-2\vert U_{\mu1} \vert^2\vert U_{\mu2} \vert^2 \sin^2 \Delta_{21}}
{(\vert U_{\mu1} \vert^2+\vert U_{\mu2} \vert^2)^2-4\vert U_{\mu1} \vert^2\vert U_{\mu2} \vert^2 \sin^2 \Delta_{21}} \right] \vert \Delta m^2_{32} \vert.
\end{eqnarray}
Again for small $\Delta_{21}$ this reduces to 
$\Delta m^2_{\mu \mu}$ given by Eq.~(\ref{eqn:dmsqmumu}).

For $\nu_\tau$ disappearance, just replace 
$\vert U_{\mu j}\vert^2 $ with $ \vert U_{\tau j} \vert^2$ in 
the equations of this appendix.


\end{document}